\pdfoutput=1
\documentclass[%
 reprint,
superscriptaddress,
 amsmath,amssymb,
 aps,
 nofootinbib,
]{revtex4-1}
\usepackage{color}
\usepackage{feynmf}
\usepackage{graphicx}
\usepackage{dcolumn}
\usepackage{bm}
\usepackage[utf8]{inputenc}
\usepackage{hyperref}
\usepackage{lipsum}
\usepackage{slashed}

\newcommand{\vev}{\varphi}

\newcommand{\sigmaMeson}{\sigma}
\newcommand{\fermion}{\Psi}
\newcommand{\antiFermion}{\bar{\Psi}}

\newcommand{\etaMeson}{\eta}
\newcommand{\Trace}[1]{\textrm{Tr}\left[#1\right]}
\newcommand{\quark}{q}

\newcommand{\Exp}[1]{\textrm{e}^{#1}}

\newcommand{\cN}{c_N} 

\newcommand{\baryon}{\Psi}

\newcommand{\chiMeson}{\chi}
\newcommand{\gAV}{g_{\textrm{AV}}}
\newcommand{\mixingAngle}{\theta}
\newcommand{\massH}{M_{H}}
\newcommand{\massS}{M_{S}}
\newcommand{\massG}{M_{G^\prime}}

\newcommand{\glueball}{G}
\newcommand{\HField}{H}
\newcommand{\SField}{S}
\newcommand{\GField}{G^\prime}

\newcommand{\QQ}[1]{Q_{#1}}

\newcommand{\SU}[2]{SU(#1)_{\textrm{#2}}}
\newcommand{\UA}{U(1)_{\textrm{A}}}
\newcommand{\Gfl}{G_{\textrm{fl}}}
\newcommand{\UU}[1]{U_{\textrm{#1}}}
\newcommand{\hc}{\textrm{h.c.}}
\newcommand{\trm}[1]{\textrm{#1}}
\newcommand{\axialParam}{\alpha}

\begin{document}


\title{Role of a four-quark and a glueball state in pion-pion and pion-nucleon scattering}

\author{Phillip Lakaschus}
\affiliation{Institute for Theoretical Physics, Goethe University, Max-von-Laue-Str.\ 1,
D-60438 Frankfurt am Main, Germany}
\author{Justin L.P.\ Mauldin}
\affiliation{Institute for Theoretical Physics, Goethe University, Max-von-Laue-Str.\ 1,
D-60438 Frankfurt am Main, Germany}
\author{Francesco Giacosa}
\affiliation{Institute of Physics, Jan Kochanowski University, PL-25406 Kielce, Poland}
\author{Dirk H.\ Rischke}
\affiliation{Institute for Theoretical Physics, Goethe University, Max-von-Laue-Str.\ 1,
D-60438 Frankfurt am Main, Germany}
\affiliation{Interdisciplinary Center for Theoretical Study and Department of Modern Physics, 
University of Science and Technology of China, Hefei, Anhui 230026, China}

\date{\today}

\begin{abstract}
 We consider the two-flavor version of the extended linear sigma
model (eLSM), which contains (pseudo)scalar and (axial-)vector
quark-antiquark mesons, 
 a scalar glueball (predominantly corresponding to $f_{0}(1710)$), as
well as the nucleon and its chiral partner. We extend this model by  the
additional light scalar meson $f_{0}(500)$, predominantly a putative four-quark state.
We investigate  various interaction
terms of  the four-quark and glueball states with the other particles, some of which preserve and
some of which explicitly break the $U(1)_{\textrm{A}}$ symmetry. We test our model by
performing a global fit to masses and decay widths of the scalar resonances
and pion-pion scattering lengths. We also discuss the influence of the scalar
four-quark state and the glueball on the baryon sector by evaluating
pion-nucleon scattering parameters. We find that the inclusion of $f_{0}(500)$
 improves the description of
pion-pion and pion-nucleon scattering lengths.
\end{abstract}

\pacs{Valid PACS appear here}
\maketitle


\section{Introduction}
\label{sec:level1}

A major  task in low-energy  hadron physics is the unified
description of masses, decays, and scattering properties (including scattering
lengths, phase shifts, etc.) of all light hadrons (both mesons and baryons)
below $\sim$2 GeV \cite{Patrignani:2016xqp}.   This problem is exceptionally
difficult, due to the large number of hadrons and the intrinsically strong
interaction between them.

Since  Quantum Chormodynamics (QCD), the fundamental theory of the strong interaction,
cannot be directly solved in  the low-energy domain, various
methods were developed to  describe mesons and baryons. The
 relativistic quark model of Refs.\ \cite{isgur1,isgur2}  solidly reproduces
properties of conventional quark-antiquark and three-quark states. Even after
many years, it still provides a useful starting point for many considerations.
Yet, the effect of mesonic quantum  corrections is not taken into account [for
 an extension in this direction, see Ref.\ \cite{vbv}] and various candidates  for
non-conventional mesons (such as glueballs, hybrids, and multiquark states)
cannot be easily  accounted for. On the other hand, numerical simulations of QCD 
 on the lattice are
now capable  of reproducing a large part of the QCD spectrum [see e.g.\ Ref.\
\cite{dudekexcited}].  Nowadays even scattering lengths can be
 computed, see e.g.\ Ref.\ \cite{dudekrho}. However, there is still a long way  to go
towards an exhaustive description of all properties of low-energy
QCD  using lattice simulations.

Another line of research has been the development of effective chiral
approaches. Some make use of quark degrees of freedom, such as the famous
Nambu--Jona-Lasinio (NJL) model \cite{njl1,njl2,njl3,njl4,njl5} [and  the related quark-meson model,
see Refs.\ \cite{quarkmeson1,quarkmeson2,quarkmeson3} and refs.\ therein]. More recently, 
Dyson-Schwinger equations  have been employed to calculate meson and baryon masses in an
approach  which starts directly from the QCD Lagrangian and  respects chiral
symmetry [for reviews, see Refs.\ \cite{ds1,ds2}].

 Another approach to describe meson properties in the low-energy domain is 
chiral perturbation theory (ChPT), see e.g.\ Refs.\ \cite{chpt1,chpt2,chpt3,chpt4,chpt5}.
 It is based on a non-linear realization of chiral symmetry and  the primary
method to study hadronic low-energy properties in a systematic and well-defined  way. 
 ChPT is  originally devised to study the interactions of the (pseudo)-Goldstone bosons
emerging from chiral symmetry breaking, i.e., for two quark flavors the pions. The description of other (and heavier) 
mesons becomes more difficult \cite{ecker1,ecker2}.

Other chiral approaches, so-called linear sigma models [see e.g.\ Refs.\
\cite{korudaz,urban,carter,Parganlija2010,Parganlija,kovacs}], are based on
the linear realization of chiral symmetry, hence contain 
 hadrons and their chiral partners on the same footing. In particular, 
within the last  ten years a
chiral model,  called extended Linear Sigma Model (eLSM), has been
developed \cite{Parganlija2010,Parganlija,JanowskiGlueball} in  an attempt to
include as many resonances as possible. The eLSM is based on both chiral
symmetry and dilatation invariance and  correctly models their  respective 
explicit, anomalous, and spontaneous breaking  mechanisms. In the mesonic sector, the eLSM contains
also (axial-)vector meson  degrees of freedom
besides  the standard (pseudo)scalar mesons.
 The lightest scalar glueball is included as a dilaton field in the Lagrangian.
Moreover, the model  was extended to include the  lightest pseudoscalar glueball
\cite{eshraim},  a vector glueball, pseudovector and excited vector mesons
\cite{sammet},  as well as excited (pseudo)scalar mesons \cite{elsmexcited}. In the
low-energy limit, the eLSM correctly reduces to ChPT \cite{elsmchpt}, thus
showing the compatibility of these two different approaches to hadronic
physics. In the baryonic sector, the eLSM was developed for two flavors
in Ref.\ \cite{Gallas} and for three flavors in Ref.\ \cite{Olbrich:2015gln} on the
basis of the mirror assignment  for the chiral partner of the nucleon \cite{detar}, 
in such a way that  chirally invariant mass terms for baryons are possible.

In general, the eLSM offers a satisfactory description of  hadronic properties
below 2 GeV [see Ref.\ \cite{Parganlija}]. In particular, the eLSM has
a clear answer  concerning the interpretation of scalar mesons
\cite{JanowskiGlueball}: the chiral partner of the pion is predominantly
$f_{0}(1370)$,  its quark structure being $\left(  \bar{u}u+\bar{d}d\right)
/\sqrt{2}$, while $f_{0}(1500)$ is predominantly $\bar{s}s.$ Then, the state
$f_{0}(1710)$ is mostly gluonic and  could be the  lightest scalar glueball
in agreement with calculations within lattice QCD \cite{mainlattice,chenlattice} and holographic QCD
\cite{Albaladejo,rebhan1,rebhan2,rebhan3} [for other  interpretations suggesting a 
mixing of quarkonium states, see Refs.\
\cite{refs1,refs2,refs3,refs4} and refs.\ therein]. The isovector state $a_{0}(1450)$ and the
isodoublet states $K_{0}^{\ast}(1430)$ complete  the nonet of
$\bar{q}q$ states, respectively. In this way, the light scalar  resonances $f_{0}(500)$, $f_{0}(980)$,
$a_{0}(980)$, and $K_{0}^{\ast}(800)$ are \textit{not} part of the eLSM.
Hence, these  resonances are not predominantly quark-antiquark states, but
something else.

There is nowadays consensus  that these resonances are 
most likely four-quark states. This still leaves 
different possibilities for the internal structure of these states: following the original
proposal  by Jaffe \cite{jaffeorig},  they could be 
bound states formed by a colored diquark (in the antisymmetric color-antitriplet and 
antisymmetric flavor-antitriplet representation) and a colored anti-diquark (in the corresponding color-triplet
and flavor-triplet representation)
\cite{maiani,thooft,fariborz,napsuciale}. In this picture, the
resonance $f_{0}(500)$ is  a $[u,d][\bar{u},\bar{d}]$ four-quark state. 
The other members of the
nonet are formed similarly using also $[us]$, $[ds]$, and the corresponding anti-diquarks.

Alternatively,  the light scalar mesons could be (loosely bound) molecular states formed from,
or unbound states in the scattering continuum of \cite{speth1,speth2,speth3,mesonicmol1,mesonicmol2,
mesonicmol3,mesonicmol4,lowscalars1,lowscalars2, pelaezrev}, 
two  color-neutral mesons.  The latter possibility
is supported by studies where they emerge as companion poles of conventional $\bar{q}q$ seed states
\cite{vbv,pennington1,pennington2,tornqvist1,tornqvist2,kleefeld} 
[for  the dynamical generation of $a_{0}(980)$
and $K_{0}^{\ast}(800)$ starting from an eLSM-inspired Lagrangian, see Refs.\
\cite{thomas,milena}]. Thus, even if the  above mentioned approaches  differ
in the interpretation of the internal structure, all agree on a predominantly four-quark
nature of  the light scalars. In this respect, the findings of the eLSM are
consistent with these results.

The scalar state $f_{0}(500)$ is particularly important since it is the
lightest state with the quantum numbers of the vacuum [for a review, see Ref.\
\cite{pelaezrev}]. It is expected to be relevant both in pion-pion,
pion-nucleon, as well as nucleon-nucleon scattering. However, this state was
not yet included in the eLSM,  although some preliminary attempts
were made in studies at non-zero density \cite{tqfinitedensity}, at nonzero
temperature \cite{tqfinitetemperature}, and of neutron-proton scattering \cite{deinet}.
Moreover, in  a comparison of the eLSM with ChPT it was
recently stressed that the  $f_{0}(500)$ is necessary for  a proper description
of  the pion-pion scattering lengths \cite{elsmchpt}.

 The main goal of the present work is the inclusion and systematic investigation 
of the light four-quark state $f_{0}(500)$ within the
eLSM. To this end,
we  consider masses, decay widths, as well pion-pion and
pion-nucleon scattering lengths, where $f_{0}(500)$ plays a
decisive role. At the same time, we shall also investigate the effects of the
glueball/dilaton field (identified with $f_{0}(1710)$) on these
 quantities. We show that the presence of both $f_{0}(500)$ and
$f_{0}(1710)$ offers a satisfactory description of
experimental results  in the meson sector and, in the baryon sector, at least an improved
description of data in comparison to models without these states. 

 This paper is organized as follows: in Sec.\ II we  couple a
four-quark field (which  is the predominant component of $f_{0}(500)$) both  to the
mesonic and  to the baryonic sector of the eLSM as an  additional chirally invariant scalar
(thus,  in order to avoid double counting one should not generate it via loop contributions 
within this approach). For  the sake of definiteness, we  will use the 
diquark-antidiquark  picture of this state in
our considerations, but this is actually  of secondary importance for the
two-flavor version of the eLSM studied here.

In the baryonic sector, we make use of the so-called mirror assignment, first
proposed in Ref.\ \cite{detar} and further studied in Refs.\
\cite{Gallas,jido1,jido2,tolos,gallaslast,sasaki1,sasaki2,wongi}: here, the
condensation of $f_{0}(500)$ (corresponding to a four-quark condensate) and of
the dilaton $f_{0}(1710)$ (corresponding to a gluon condensate) 
 contribute to the baryonic mass terms of the mirror model  in addition to
the condensation of the standard quark-antiquark meson $f_{0}(1370)$.

Then, in Sec.\ III we present our results for masses, decay  widths, and pion-pion
as well as pion-nucleon scattering lengths and volumina.
Here,  one observes that $f_{0}(500)$ is crucial to obtain a correct
description of data  for the pion-pion scattering length $a_{0}^{0}$ and at least an improved
description of data for the
pion-nucleon scattering length $a_{0}^{(+)}.$ Moreover, a detailed study of
mixing between a bare four-quark state, a bare quark-antiquark state, and a
bare glueball confirms that the field $f_{0}(500)$ is mostly  a four-quark state,
$f_{0}(1370)$ is mostly quarkonium, and $f_{0}(1710)$ is mostly gluonic.
 We conclude this work with a summary and a discussion of the results in Sec.\ IV.
Details of the calculations are deferred to the Appendix. We use natural units $\hbar= c = 1$; the
convention for the metric tensor of flat Minkowski space-time is $g_{\mu \nu} = \textrm{diag} (+,-,-,-)$.

\section{The Model}
\label{sec:model}

\subsection{Quarkonium multiplets}
\label{sec:quarkonia}

For  an arbitrary number of flavors we can arrange (pseudo)scalar quarkonium fields  into 
multiplets using the current
\begin{equation}
\Phi^{ij} \equiv \bar{q}_{\trm{R}}^i q_\trm{L}^j \, ,
\end{equation}
which is a matrix in flavor space and where the color indices are implicitly contracted. This is a so-called 
\textit{heterochiral} \cite{Giacosa:2017pos} scalar because it transforms under  the
global chiral symmetry $\Gfl \times \UA = \SU{N_\textrm{f}}{L} \times \SU{N_\textrm{f}}{R} \times \UA$ as 
\begin{equation}
\Phi \longrightarrow \Exp{-2 i \axialParam} \UU{L} \, \Phi \, \UU{R}^\dag \, ,
\end{equation}
where $\UU{L,R}$ are  $\SU{N_\textrm{f}}{L,R}$ transformations 
and $\axialParam$ is the  parameter of the $\UA$ transformation.
For two flavors $\Phi$ can be written as
\begin{equation}
\Phi = \sum_{a=0}^{3} \Phi^a t^a = (\sigma + i \eta) t^0 + (\vec{a}_0 + i \vec{\pi})\cdot \vec{t} \, ,
\end{equation}
where $t^a = \tau^a/2$ are the generators  of $U(2)$, namely  half the Pauli matrices for $a=1, 2, 3$, 
and  half the unit matrix for $a = 0$. Similarly, we define the right- and left-handed vector currents,
\begin{equation}
R_\mu^{ij} \equiv \bar{q}_{\trm{R}}^i \gamma_\mu q_\trm{R}^j \, , \quad L_\mu^{ij} 
\equiv \bar{q}_{\trm{L}}^i \gamma_\mu q_\trm{L}^j \, .
\end{equation}
These are \textit{homochiral} multiplets, i.e., they transform under the 
 chiral symmetry $\Gfl \times \UA$ as
\begin{align}
L_\mu \longrightarrow \UU{L} \, L_\mu \, \UU{L}^\dag \, , \quad R_\mu \longrightarrow \UU{R} \, R_\mu \, \UU{R}^\dag \, .
\end{align}
For two flavors the (axial-)vector fields are contained in the right- and left-handed meson matrices
\begin{align}
R_\mu &= \sum_{a=0}^{3} R_\mu^a t^a = (\omega_\mu  - f_{1,\mu} )t^0 
+ (\vec{\rho}_\mu - \vec{a}_{1,\mu}  )\cdot \vec{t} \, , \\
L_\mu &= \sum_{a=0}^{3} L_\mu^a t^a = (\omega_\mu  + f_{1,\mu}) t^0 
+ (\vec{\rho}_\mu  + \vec{a}_{1,\mu}) \cdot \vec{t} \, .
\end{align}
These non-exotic multiplets lead to the well-known chirally invariant Lagrangian \cite{Gallas, Divotgey, Parganlija} 
\begin{widetext}
\begin{align}\label{eq:mesonLagrangian}
\mathcal{L}_{\textrm{eLSM}} &= \textrm{Tr} \left[ (D_{\mu} \Phi)^{\dag}(D^{\mu} \Phi) 
- \mu^{2} \dfrac{G^2}{G_0^2}  \Phi^{\dag} \Phi - \lambda_{2} (\Phi^{\dag} \Phi)^{2} \right] 
- \lambda_{1} (\textrm{Tr}[\Phi^{\dag} \Phi])^{2} + c(\textrm{det} \, \Phi^\dag + \hc) 
+ h_{0} \textrm{Tr}[\Phi^{\dag} + \Phi]  \notag\\
&- \dfrac{1}{4} \textrm{Tr} \left[(L^{\mu \nu})^{2} + (R^{\mu \nu})^{2} \right] 
+ \dfrac{m_{1}^{2}}{2}\dfrac{G^2}{G_0^2} \textrm{Tr} \left[(L^{\mu})^{2} + (R^{\mu})^{2} \right] 
+ \dfrac{h_{1}}{2} \textrm{Tr}[\Phi^{\dag} \Phi] \, \textrm{Tr} \left[(L^{\mu})^{2} + (R^{\mu})^{2} \right] \notag\\
& + h_{2} \textrm{Tr} \left[ \Phi^{\dag} L_{\mu} L^{\mu} \Phi + \Phi R_{\mu} R^{\mu} \Phi^{\dag} \right] 
+ 2 h_{3} \textrm{Tr} \left[\Phi  R_{\mu} \Phi^{\dag} L^{\mu} \right] 
+ i \dfrac{g_{2}}{2} \bigg( \textrm{Tr} \big[ L_{\mu \nu} [L^{\mu}, L^{\nu}] \big] 
+ \textrm{Tr} \big[ R_{\mu \nu} [R^{\mu}, R^{\nu}] \big] \bigg) \notag \\
&+ \mathcal{L}_{g_3, g_4, g_5, g_6} -V_{\textrm{dil}} (G) \, ,
\end{align}
\end{widetext}
where $D_\mu = \partial_\mu + i g_1 \left( \Phi R_\mu - L_\mu \Phi \right)$ and
\begin{equation}\label{eq:dilatonPotential}
V_{\textrm{dil}} (G) = \dfrac{1}{4} \dfrac{m_G^2}{\Lambda_{\textrm{dil}}^2} G^4 
\left( \textrm{ln} \left| \dfrac{G}{\Lambda_{\textrm{dil}}}  \right| - \dfrac{1}{4} \right) \,
\end{equation}
is the dilaton potential, responsible for the breaking of dilatation symmetry (trace anomaly) 
\cite{Salomone, Migdal:1982jp, Gomm:1984zq, Gomm2}. The scalar glueball with mass $m_G$ emerges 
upon the shift $G \rightarrow G_0 + G$. The scalar glueball represents a color-neutral gluonic bound state. 
Since it is not composed of any quarks of any flavor, it is trivially a chiral singlet.

The Lagrangian $ \mathcal{L}_{g_3, g_4, g_5, g_6}$ describes (axial-)vector meson interactions,  see 
Appendix of Ref.\ \cite{Gallas}. After
chiral symmetry breaking and a shift of the axial-vector fields to eliminate bilinear mixing terms between
the latter and the pions, derivatively coupled four-pion interactions emerge. Therefore, in principle the coupling constants
$g_3, g_4, g_5, g_6$ have an influence on the pion-pion scattering lengths. However, it was shown in 
Ref.\ \cite{elsmchpt} that
varying the values for these coupling constants between $\pm 100$ (i.e., a range of values that is in agreement with
ChPT and moreover appears 
to be of a natural order of magnitude) the change in the pion-pion scattering lengths is only of the order of
a few percent. Therefore,
we will neglect these coupling terms in the following.

\subsection{Four-quark multiplets}
\label{sec:tetraquarks}

There are several ways to  incorporate four-quark states  into a chiral model. 
Here, we will follow the approach of Ref.\ \cite{Giacosa:2006tf}.  We start with the three-flavor case and 
 then reduce it to two flavors.  There are several reasons for choosing this 
approach:\\
\begin{enumerate}
\item Writing down all relevant terms  for three flavors  facilitates an extension of
the current work  to the case of three flavors.
\item We will be able to compare our approach with other ones, e.g.\ 
 those of Refs.\ \cite{Pisarski:2016ukx, FariborzToyModel}.
\item It is  easier to see which terms are large-$N_c$ dominant, enabling us 
to choose only the most relevant terms.
\item  For three flavors a four-quark nonet  has the same chiral structure as the 
quarkonium nonet, while for $N_\textrm{f} \ne 3$ the four-quark multiplet will have a different chiral structure, 
which makes three-flavor models somewhat special when considering four-quark states \cite{FariborzThreeFlavors}.
\end{enumerate}

Here, we use a diquark-antidiquark picture as a concrete
framework to construct the multiplet of light scalars and to couple them to
conventional mesons. However, it is also possible to construct the same terms
in the meson-meson molecular picture \cite{FariborzToyModel}. In the two-flavor eLSM both
approaches yield the same effective Lagrangian at leading order in the
large-$N_{c}$ expansion. In general, when we refer to four-quark states we
understand both diquark-antidiquark as well as meson-meson components; even if
the latter are expected to be dominant \cite{pelaezrev}, an admixture of the
former configurations is possible.

For $N_\textrm{f} = 3$ the right- and left-handed diquark  fields are defined as
\begin{align}\label{eq:leftRighthandedDiquarks}
L_{cC} &= \epsilon_{cab} \epsilon_{CAB} \, \quark^\trm{T}_{aA} C \mathcal{P}_\trm{L} \quark_{bB} \, , \\
R_{cC} &= \epsilon_{cab} \epsilon_{CAB} \, \quark^\trm{T}_{aA} C \mathcal{P}_\trm{R} \quark_{bB} \, ,
\end{align}
with  the charge-conjugation matrix $C=i\gamma^2\gamma^0$ (in the Dirac representation), where 
the flavor indices are in small letters and color indices in capital letters. For  the sake of
simplicity we will drop the color indices in 
the following. In the next step, we construct the right and left-handed diquark matrices
\begin{align}\label{eq:diquarkMatrix}
D_{\trm{L},ab} = (A_c)_{ab} L_c  \, , \quad D_{\trm{R},ab} = (A_c)_{ab} R_c  \, ,
\end{align}
where $(A_c)_{ab} \equiv \epsilon_{cab}$.
They transform as $\Exp{-2i \alpha} \UU{L} D_{\trm{L}} \UU{L}^\textrm{T}$ and 
$\Exp{2i \alpha}\UU{R} D_{\trm{R}} \UU{R}^\textrm{T}$ under $\Gfl \times \UA$, respectively. 
Under parity $D_\trm{L/R}$ transforms  into $-D_\trm{R/L}$ and 
under charge conjugation  into $-D_\trm{R/L}^\dag$, 
such that we obtain diquark matrices with well defined parity  via the linear combinations
\begin{equation}\label{eq:(pseudo)scalarDiquarkMatrices}
    D = \dfrac{D_\trm{R} - D_\trm{L}}{\sqrt{2}} \, , \quad \tilde{D} = \dfrac{D_\trm{R} + D_\trm{L}}{\sqrt{2}} \, .
\end{equation}
These are composed of scalar and pseudoscalar diquarks, defined as
\begin{align}
S_{cC} &= \dfrac{1}{\sqrt{2}} \epsilon_{cab} \epsilon_{CAB} \, \quark^T_{aA} C \gamma^5 \quark_{bB} \, , \\
P_{cC} &= \dfrac{1}{\sqrt{2}} \epsilon_{cab} \epsilon_{CAB} \, \quark^T_{aA} C \quark_{bB} \, ,
\end{align}
such that (again suppressing color indices)
\begin{equation}
S_c = \dfrac{R_c - L_c}{\sqrt{2}} \, , \quad P_c = \dfrac{R_c + L_c}{\sqrt{2}} \, .
\end{equation}
In the following we will  only be interested in scalar diquarks, assuming that the 
pseudoscalar ones are not relevant for low-energy hadron phenomenology. 
A strong attraction between two quarks in a color antitriplet ($\overline{3}_{\mathrm{C}}$), a flavor antitriplet
($\overline{3}_{\mathrm{F}}$) and  spin-zero configuration \cite{jaffeorig}  is obtained in studies
based on one-gluon exchange \cite{onegluon}, instantons
\cite{inst1,inst2}, the NJL model \cite{weise}, and
Dyson-Schwinger equation (DSE) \cite{maris}. In these studies it is also shown that the pseudoscalar diquark 
turns out to be considerably heavier. 

Considering only scalar diquarks we can therefore construct a scalar  four-quark nonet by 
\begin{equation}\label{eq:tetraquarkMatrix}
T_{ab} = S_a^\dag S_b \, ,
\end{equation}
which explicitly reads
\begin{align}\label{eq:explicitTetraquarkMatrix}
T &= \begin{pmatrix}
[\bar{d},\bar{s}][d,s] & [\bar{d},\bar{s}][u,s] & [\bar{d},\bar{s}][u,d] \\
[\bar{u},\bar{s}][d,s] & [\bar{u},\bar{s}][u,s] & [\bar{u},\bar{s}][u,d] \\
[\bar{u},\bar{d}][d,s] & [\bar{u},\bar{d}][u,s] & [\bar{u},\bar{d}][u,d] 
\end{pmatrix} \, \notag\\
& = \begin{pmatrix}
\sqrt{\dfrac{1}{2}} (\chiMeson_\textrm{s} - a_0^0) & -a_0^+ &  K_0^{*+} \\
-a_0^- & \sqrt{\dfrac{1}{2}} (\chi_\textrm{s} + a_0^0) & - K_0^{*0} \\
 K_0^{*-} & - \bar{K}_0^{*0} & \chiMeson
\end{pmatrix} \, ,
\end{align}
where $a_0$ and $ K_0^{*}$ are identified with $a_0(980)$ and $ K_0^{*}(800)$, 
and  admixtures of 
$\chiMeson$ and $\chiMeson_\textrm{s}$  are assigned to the physical states $f_0(500)$ and $f_0(980)$.
Now we are able to construct chirally invariant interaction terms that couple scalar  four-quark states 
to scalar quarkonia:
\begin{align}\label{eq:allScalarTetraquarkInteractions}
\mathcal{L}_{\textrm{T}\Phi\Phi} &= g_\chiMeson^{(1)} \dfrac{G}{G_0} \Trace{D_\trm{R} \Phi^\trm{T} D_\trm{L}^\dag \Phi 
+ D_\trm{L} \Phi^* D_\trm{R}^\dag \Phi^\dag } \notag\\
&+g_\chiMeson^{(2)} \dfrac{G}{G_0} \Trace{D_\trm{R} D_\trm{R}^\dag \Phi^\dag \Phi 
+ D_\trm{L} D_\trm{L}^\dag \Phi \Phi^\dag} \notag\\
&+ g_\chiMeson^{(3)} \dfrac{G}{G_0} \Trace{D_\trm{R} D_\trm{R}^\dag 
+ D_\trm{L} D_\trm{L}^\dag} \Trace{\Phi \Phi^\dag} \, .
\end{align}
In each term there is a diquark and an anti-diquark such that the expression becomes color neutral. Furthermore, all 
terms are also invariant under parity, charge conjugation, and $\UA$  transformations. In order to keep our effective 
model as simple as possible,  in the following we will only consider the first term 
 in Eq.\ \eqref{eq:allScalarTetraquarkInteractions}, which is  the leading one 
in the large-$N_\trm{c}$ expansion,
\begin{align}\label{eq:scalarTetraquarkTerm}
\mathcal{L}_{\textrm{T}\Phi\Phi} &= -\dfrac{g_\chiMeson^{(1)}}{2} \dfrac{G}{G_0} \Trace{D \Phi^\trm{T} D^\dag \Phi 
+ D \Phi^* D^\dag \Phi^\dag } + \ldots \notag\\
&= -\dfrac{g_\chiMeson^{(1)}}{2} \dfrac{G}{G_0} T_{ab} \Trace{A_b \Phi^\trm{T} A_a^\trm{T} \Phi 
+ A_b \Phi^* A_a^\trm{T} \Phi^\dag } + \ldots \, ,
\end{align}
where we used Eqs.\ \eqref{eq:(pseudo)scalarDiquarkMatrices} and \eqref{eq:tetraquarkMatrix} and neglected all terms 
containing pseudoscalar diquarks.\footnote{ We note that this term is precisely the 
same as the one  in Eq.\ (16)  of Ref.\ \cite{FariborzToyModel}. However, 
in  that work the interaction term is 
constructed by using the  four-quark matrix analogue of $\Phi$, which transforms in the same manner as 
$\Phi$ except  under $\UA$ transformations. 
Although both approaches yield the same interaction term as above, 
the other two interaction terms of Eq.\ \eqref{eq:allScalarTetraquarkInteractions} are only found using our approach.} 

 For two flavors, Eq.\ \eqref{eq:scalarTetraquarkTerm} reduces to
\begin{align}\label{eq:explicitScalarTetraquarkInteraction}
\mathcal{L}_{\textrm{T}\Phi\Phi} &=  2 g_\chiMeson \dfrac{G}{G_0} \chiMeson 
\left( \textrm{det} \, \Phi + \hc \right) \notag \\
&=  g_\chiMeson \dfrac{G}{G_0} \chiMeson \left( \sigma^2 + \vec{\pi}^{\,2} - \etaMeson^2 
- \vec{a}_0^{\,2} \right) \, ,
\end{align}
where we  abbreviated $g_\chiMeson^{(1)} \equiv -2 g_\chiMeson$ and only considered 
$T_{33} \equiv \chiMeson$ because this is the only  four-quark state that exists  for 
two flavors.
This term is similar to the determinant term in Eq.\ \eqref{eq:mesonLagrangian}, which models the $\UA$ anomaly. 
Since the chiral condensate also induces a condensate of the scalar  four-quark state, 
Eq. \eqref{eq:scalarTetraquarkTerm} generates a contribution to the masses of 
$\sigma$ and $\pi$ and, of the same magnitude but with opposite sign, to those of $\eta$ and $a_0$.

Very similar terms are obtained for the coupling to (axial-)vector quarkonia:
\begin{align}\label{eq:allVectorTetraquarkInteractions}
\mathcal{L}_{\textrm{T}-\trm{AV}} &= \gAV^{(1)} \dfrac{G}{G_0} \Trace{D_\trm{R} R_\mu^\trm{T} D_\trm{R}^\dag R^\mu 
+ D_\trm{L} L_\mu^\trm{T} D_\trm{L}^\dag L^\mu} \notag\\
&+\gAV^{(2)} \dfrac{G}{G_0} \Trace{D_\trm{R} D_\trm{R}^\dag R_\mu^\dag R^\mu 
+ D_\trm{L} D_\trm{L}^\dag L_\mu^\dag L^\mu } \notag\\
&+ \gAV^{(3)} \dfrac{G}{G_0} \Trace{D_\trm{R} D_\trm{R}^\dag + D_\trm{L} D_\trm{L}^\dag} \Trace{R_\mu^\dag R^\mu 
+ L_\mu^\dag L^\mu} \, .
\end{align}
Again, we are only interested in the leading-order term  at large-$N_\trm{c}$. 
 Neglecting the pseudoscalar diquark $\tilde{D}$ we obtain:
\begin{align}
\mathcal{L}_{\textrm{T}-\trm{AV}} &= \dfrac{\gAV^{(1)}}{2} \dfrac{G}{G_0}\Trace{D R_\mu^\trm{T} D^\dag R^\mu 
+ D L_\mu^\trm{T} D^\dag L^\mu} \notag \\
&= \dfrac{\gAV^{(1)}}{2} \dfrac{G}{G_0} T_{ab} \Trace{A_b R_\mu^\trm{T} A_a^\trm{T} R^\mu + A_b L_\mu^\trm{T} A_a^\trm{T} L^\mu} \, .
\end{align}
\noindent For two flavors, this term reduces to:
\begin{align}\label{eq:gAVinteraction}
\mathcal{L}_{\textrm{T}-\trm{AV}}&= -2\, \gAV \, \dfrac{G}{G_0} \chiMeson (\textrm{det} \, R_\mu 
+ \textrm{det} \, L_\mu)    \notag \\
&=   \gAV \, \dfrac{G}{G_0} \chiMeson \left( \vec{\rho}_\mu^{\,2} 
+ \vec{a}_{1,\mu}^{\,2} - \omega_\mu^2 - f_{1,\mu}^2 \right) \, ,
\end{align}

\noindent where we abbreviated $ \gAV^{(1)} \equiv - 2 \gAV$.
Interestingly, this term looks structurally similar to Eq.\ \eqref{eq:explicitScalarTetraquarkInteraction}. 
It generates a  contribution to the masses of the isoscalar (axial-)vector mesons and, of the
same magnitude but with opposite sign, to those of the isovector (axial-)vector mesons. 

Introducing also a kinetic and mass term for the scalar tetraquark, we find the complete two-flavor 
 four-quark Lagrangian to be
\begin{align}\label{eq:completeTetraquark}
\mathcal{L}_{\chiMeson-\textrm{int}} &= \dfrac{1}{2}  \partial_\mu \chiMeson \partial^\mu \chiMeson 
- \dfrac{1}{2} m_\chiMeson^2 \dfrac{G^2}{G_0^2} \chiMeson^2 \notag \\
& +  g_\chiMeson \dfrac{G}{G_0} \chiMeson \left( \sigma^2 + \vec{\pi}^2 
- \etaMeson^2 - \vec{a}_0^2 \right)  \notag \\
&+ \gAV \dfrac{G}{G_0} \chiMeson \left( \vec{\rho}_\mu^{\,2} + \vec{a}_{1,\mu}^2 
- \omega_\mu^2 - f_{1,\mu}^2 \right) \, .
\end{align}
From this Lagrangian and Eq.\ \eqref{eq:mesonLagrangian} we can derive masses and 
decay widths as well as the pion-pion scattering lengths. 

\subsection{Masses of the scalar-isoscalar resonances}
\label{sec:glueball}

 The terms in the Lagrangian which, upon condensation of $\sigma, G,$ and $\chi$, 
give rise to the mass matrix for the scalar-isoscalar resonances are
\begin{align}\label{eq:completeGlueballInteraction}
&  -\dfrac{1}{2}\mu^2 \dfrac{G^2}{G_0^2} \sigma^2  + \dfrac{c}{2} \sigma^2 
-\dfrac{1}{4}\left( \lambda_1 + \dfrac{\lambda_2}{2} \right)  \sigma^4 \notag\\
&-\dfrac{1}{2} m_\chiMeson^2 \dfrac{G^2}{G_0^2} \chiMeson^2 + g_\chiMeson \dfrac{G}{G_0} \chiMeson \sigma^2 \notag \\
& - \dfrac{1}{4} \dfrac{m_G^2}{\Lambda_{\textrm{dil}}^2} G^4 
\left( \textrm{ln} \left| \dfrac{G}{\Lambda_{\textrm{dil}}}  \right| - \dfrac{1}{4} \right)\,.
\end{align}
We perform a shift of the scalar-isoscalar fields by their respective vacuum expectation values,
$\sigma \rightarrow \varphi + \sigma$, 
$G \rightarrow G_0 + G$, $\chi \rightarrow \chi_0 + \chi$. This
leads to mass terms for these three fields, which can be compactly written in matrix form as
\begin{widetext}
\begin{equation}\label{eq:massMatrixNeW}
V_{\textrm{mass}}(\chiMeson, \sigmaMeson,G) = \dfrac{1}{2} 
(\chiMeson, \, \sigmaMeson, \, G)
\begin{pmatrix}
m_{\chiMeson}^2 & -2 g_{\chiMeson} \vev 
& g_\chiMeson \dfrac{\vev^2}{G_0}  \\
-2 g_{\chiMeson} \vev &  m_{\sigmaMeson}^2 & \dfrac{2 \vev}{G_0} (\mu^2 - g_\chiMeson \chiMeson_0) \\
g_\chiMeson \dfrac{\vev^2}{G_0} 
& \dfrac{2 \vev}{G_0} (\mu^2 - g_\chiMeson \chiMeson_0) & M_G^2
\end{pmatrix} \begin{pmatrix}
\chiMeson \\
\sigmaMeson \\
G
\end{pmatrix} \equiv \, \dfrac{1}{2} 
(\chiMeson, \, \sigmaMeson, \, G) \; M \begin{pmatrix}
\chiMeson \\
\sigmaMeson \\
G
\end{pmatrix}\, ,
\end{equation}
\end{widetext}
 where $m_\sigmaMeson^2$ is given in Eq.\ (\ref{eq:sigmamass}) and
\begin{align}
M_G^2 & = \dfrac{\mu^2 \varphi^2 + m_\chi^2 \chi_0^2}{G_0^2} + m_G^2 \dfrac{G_0^2}{\Lambda_{\textrm{dil}}^2}
\left( 1+ 3 \ln \left| \dfrac{G_0}{\Lambda_{\textrm{dil}}} \right|  \right)\,.
\end{align}
Note that the (13) and (31) elements of $M$ were simplified using the condition that $\chi_0$ is an extremum
of the potential energy density.
The real, symmetric mass matrix $M$ can be diagonalized by an orthogonal transformation $O$,
$O^{T} M O = M_{\textrm{diag}}$,  
\begin{equation}
M_{\textrm{diag}} = \begin{pmatrix}
M_{\HField}^2 & 0 & 0  \\
0 & M_{\SField}^2 & 0\\
0 & 0 & M_{\GField}^2
\end{pmatrix} \, ,
\end{equation}
 where the eigenvalues $M_{\HField}^2$, $M_{\SField}^2$, and $M_{\GField}^2$ 
correspond to the (squared) masses of the physical fields $f_0(500)$, $f_0(1370)$, and $f_0(1710)$.
These fields are linear combinations of the unphysical fields $\chi$, $\sigma$, and $G$, given by
\begin{equation}\label{eq:threeRotation}
\begin{pmatrix}
f_0(500) \\
f_0(1370) \\
f_0(1710)
\end{pmatrix}
= 
\begin{pmatrix}
\HField \\
\SField \\
\GField
\end{pmatrix}
=
 O^T
\begin{pmatrix}
\chiMeson \\
\sigmaMeson \\
\glueball
\end{pmatrix} \, .
\end{equation}

\subsection{Baryons}
\label{sec:baryonSector}

 Baryons are implemented in the eLSM in the so-called 
mirror assignment \cite{detar, Jido:2001nt,Gallas}.
One  introduces two baryon doublets, $\baryon_1$ and $\baryon_2$, where $\baryon_1$ has 
positive parity and $\baryon_2$ is its chiral partner  with negative parity. In the mirror assignment 
these fields transform  under chiral transformations as
\begin{equation}
\baryon_{1,\mathrm{R/L}} \rightarrow \UU{R/L} \baryon_{1, \mathrm{R/L}}\, , \quad 
\baryon_{2,\mathrm{R/L}} \rightarrow \UU{L/R} \baryon_{2, \mathrm{R/L}}\, ,
\end{equation}
 i.e., $\baryon_{1,\mathrm{R/L}}$ transforms like a vector under $SU(N_\textrm{f})_{\mathrm{R/L}}$, 
as expected, while
$\baryon_{2,\mathrm{R/L}}$ transforms in a mirror way,  like a vector under $SU(N_\textrm{f})_{\mathrm{L/R}}$. 
Both fields are singlets under
$U(1)_\textrm{A}$ transformations.

 The mirror assignment allows for the existence of a new chirally invariant mass term, 
which contributes to the baryon masses in a different manner than the chiral condensate. Thus, baryons can have
non-zero masses even when the chiral condensate vanishes. Demanding dilatation invariance, the
new mass term must necessarily arise from coupling the baryons to the four-quark field $\chi$,
\begin{align}\label{eq:gallasTetraquarkNucleon}
    - a \chiMeson \left(\antiFermion_{1\mathrm{L}} \fermion_{2\mathrm{R}} 
    - \antiFermion_{1\mathrm{R}} \fermion_{2\mathrm{L}} + \hc \right) \, ,
\end{align}
and/or the glueball field $G$,
\begin{equation}\label{eq:glueballInt}
    - b G \left(\antiFermion_{1\mathrm{L}} \fermion_{2\mathrm{R}} 
    - \antiFermion_{1\mathrm{R}} \fermion_{2\mathrm{L}} + \hc \right) \, ,
\end{equation}
and subsequent condensation of $\chi$ and $G$. 
Both $G$ and $\chiMeson$ are chiral singlets as shown in App.\ \ref{app:tetraquarkBaryon}.

Furthermore, it is possible to introduce another interaction term that violates $\UA$ symmetry but gives a 
contribution to the nucleon mass as well. This term is given by
\begin{align}
    \mathcal{L}_{\textrm{anom.}} &= - \cN \left( \textrm{det} \, \Phi + \hc \right)  
    \left(\antiFermion_{1\mathrm{L}} \fermion_{2\mathrm{R}} - \antiFermion_{1\mathrm{R}} \fermion_{2\mathrm{L}} 
    + \hc \right) \notag \\
    & = - \dfrac{\cN}{2} \left(\sigmaMeson^2 + \vec{\pi}^2 - \etaMeson^2 - \vec{a}_0^2 \right) \notag \\
    & \times \left(\antiFermion_{1\mathrm{L}} \fermion_{2\mathrm{R}} - \antiFermion_{1\mathrm{R}} \fermion_{2\mathrm{L}}
     + \hc \right) \, .
\end{align}
 Such an anomalous term (which also breaks the dilatation symmetry) yields a four-point vertex
that has not been considered before in this model, see also the last diagram of Fig.\ \ref{fig:treeLevelDiagrams} (found in App. \ref{app:piNScatFormulas}).

 The terms in the baryon Lagrangian which are relevant for pion-nucleon scattering are then
\begin{widetext}
\begin{align}\label{eq:baryonLagrangian}
\mathcal{L}_{\textrm{eLSM}}^{\textrm{bar}} &= \antiFermion_{1\mathrm{L}} i \gamma_{\mu} 
D^{\mu}_{1\mathrm{L}} \fermion_{1\mathrm{L}} 
+ \antiFermion_{1\mathrm{R}} i \gamma_{\mu} D^{\mu}_{1\mathrm{R}} \fermion_{1\mathrm{R}} 
+ \antiFermion_{2\mathrm{L}} i \gamma_{\mu} D^{\mu}_{2\mathrm{R}} \fermion_{2\mathrm{L}} 
+ \antiFermion_{2\mathrm{R}} i \gamma_{\mu} D^{\mu}_{2\mathrm{L}} \fermion_{2\mathrm{R}} \notag\\ 
&- \hat{g}_{1} (\antiFermion_{1\mathrm{L}} \Phi \fermion_{1\mathrm{R}} + \hc) 
- \hat{g}_{2} (\antiFermion_{2\mathrm{L}} \Phi^{\dag} \fermion_{2\mathrm{R}} + \hc) 
- \left[a \chiMeson + b G + \dfrac{\cN}{2} (\sigmaMeson^2 + \vec{\pi}^2)\right] 
\left(\antiFermion_{1\mathrm{L}} \fermion_{2\mathrm{R}} - \antiFermion_{1\mathrm{R}} \fermion_{2\mathrm{L}} 
+ \hc \right) + \ldots\;,
\end{align}	
\end{widetext}
where $D_{1/2,\mathrm{R}}^\mu = \partial^\mu - i c_{1/2} R^\mu$, 
$D_{1/2,\mathrm{L}}^\mu = \partial^\mu - i c_{1/2} L^\mu$.

Upon condensation of $\chiMeson$, $G$, and $\sigmaMeson$ a baryonic mass term is generated
\begin{equation}\label{eq:m0expression}
    m_0 \equiv a \chiMeson_0 + b G_0 + \dfrac{\cN \vev^2}{2} \, .
\end{equation}
The mass term  mixes $\Psi_1$ and $\Psi_2$, so that the physical fields are obtained by a unitary 
 transformation,
\begin{equation}
\begin{pmatrix}
N \\
N^{*}
\end{pmatrix}
= \dfrac{1}{\sqrt{2 \textrm{cosh} \, \delta}} 
\begin{pmatrix}
e^{\delta /2}  &   \gamma_{5} e^{-\delta /2}\\
\gamma_{5} e^{-\delta /2}  &  - e^{\delta /2}
\end{pmatrix} 
\begin{pmatrix}
\Psi_{1} \\
\Psi_{2}
\end{pmatrix} \, ,
\end{equation}
where $\delta$ is the mixing parameter.

We recall that the quantity $m_{0}$ does not represent the
baryon mass in the chiral limit (i.e., when the bare quark masses are set to
zero), but represents the chirally invariant contribution to the nucleon mass,
which is the same for both the nucleon $N(939)$ and its chiral partner
$N(1535)$. As Eq.\ (\ref{eq:m0expression}) shows, in our case $m_{0}$ consists of three
contributions: the scalar tetraquark condensate $\chi_{0}$, the dilaton
condensate $G_{0}$, and also the chiral condensate $\varphi$. The difference
in mass between $N$ and $N^{\ast}(1535)$ is, on the contrary, solely
proportional to the chiral condensate $\varphi.$ If we consider the limit
$\varphi\rightarrow0$ (no SSB) one obtains $m_{0}\rightarrow bG_{0}\neq0,$
hence without SSB the nucleon and its chiral partner would have an
identical nonzero mass. [Note that $\chi_{0}$ vanishes also when $\varphi
\rightarrow0$ since it is proportional to $\varphi^{2}$ \cite{tqfinitetemperature}].

\section{Results}
\label{sec:results}

In this section we  first perform a global fit of the parameters in the meson sector.
Here we consider two different scenarios: first we neglect the scalar glueball and investigate the 
 mixing of the scalar four-quark with the quarkonium state only. Then, we present the 
results  for the full 
three-scalar mixing problem, which includes the scalar glueball,  the four-quark, and the quarkonium state.
 This allows us to estimate the importance of the scalar glueball for the calculation of the 
decay widths of the 
scalar-isoscalars and the pion-pion scattering lengths.  Subsequently, 
we will take the results from the global fit of the 
meson sector  and calculate pion-nucleon scattering parameters.

\subsection{Global fit in the meson sector}\label{sec:mesonResults}

The assignment of our effective hadronic 
degrees of freedom is given in Tab.\ \ref{tab:assignments}. For this assignment the experimental data for the decay 
widths and the pion-pion scattering lengths  are given in Tab.\ \ref{tab:decay}. 
For the mass of the pion we use the isospin-averaged mass, 
$m_\pi = 138 \, \textrm{MeV}$.  For the mass of $a_1$ we choose 1277 \textrm{MeV}, 
which is slightly above the upper error band of the PDG data. The reason for this choice is the mass 
 splitting between 
the isoscalar and the isovector (axial-)vector mesons generated by the  coupling
between the four-quark state and the \mbox{(axial-)vector} mesons in Eq.\ \eqref{eq:gAVinteraction},
\begin{equation}
   m_\rho^2 - m_\omega^2 = m_{a_1}^2 - m_{f_1}^2 = 4 \gAV \chiMeson_0 \, ,
\end{equation}
see also App.\ \ref{app:masses} and \ref{app:parameters}. Since the masses of $\rho$, $\omega$, and $f_{1}$ are known to very good precision, see Tab.\ \ref{tab:assignments},
we are forced to increase the theoretical value for the 
mass of $a_1$ such that the mass splitting between that state and $f_1$
is of the same order as that between $\rho$ and $\omega$.

Furthermore,  the physical $\eta$ meson contains a considerable $\bar{s}s$ admixture, which,
in a pure two-flavor scenario,  has to be
eliminated by a rotation in the $\eta-\eta'$ sector. 
The result is the value $m_{\eta} = 755 \, \textrm{MeV}$ for the mass of the purely non-strange 
$\etaMeson$ meson \cite{Parganlija}. 

\begin{table}[ht]
  \centering
  \begin{tabular}{ |c|c|c| }
  \hline 
  \textbf{Field} &  \textbf{Assignment} & \textbf{Masses}\\ \hline\hline
  $\HField$ & $f_0(500)$ & $475 \pm 75$ MeV\\ \hline
  $\SField$ & $f_0(1370)$ & $1350 \pm 150$ MeV\\ \hline
  $\GField$ & $f_0(1710)$ & $1723 \pm 5$ MeV\\ \hline
  $a_0$ & $a_0(1450)$  & $1474 \pm 19$ MeV\\\hline
  $a_1$ & $a_1(1260)$ & $1230 \pm 40$ MeV\\\hline
  $\rho$  & $\rho(770)$  & $775.26 \pm 0.25$ MeV\\\hline
  $f_1$  & $f_1(1285)$  & $1281.9 \pm 0.5$ MeV\\\hline
  $\omega$  & $\omega(782)$  & $782.65 \pm 0.12$ MeV\\\hline
\end{tabular}\caption{The masses of the fields  as given by the 
PDG \cite{Patrignani:2016xqp}.}\label{tab:assignments} 
\end{table}

\begin{table}[ht]
  \centering
  \begin{tabular}{ |c|c|c| }
  \hline 
  \textbf{Observable} & \textbf{Experimental Data}\\ \hline\hline
  $\Gamma_{\HField \rightarrow \pi \pi}$ & $550 \pm 150$ MeV\\ \hline
  $\Gamma_{\SField \rightarrow \pi \pi}$ & $350 \pm 150$ MeV\\ \hline
  $\Gamma_{\GField \rightarrow \pi \pi}$ & $29.3 \pm 6.5$ MeV\\ \hline
  $m_\pi a_0^0$& $0.218 \pm 0.02$ \\ \hline
  $m_\pi a_0^2$ & $-0.046 \pm 0.013$ \\ \hline
\end{tabular}
\caption{The decay widths $\Gamma_{\HField \rightarrow \pi \pi}$ and 
$\Gamma_{\GField \rightarrow \pi \pi}$  as given by the PDG \cite{Patrignani:2016xqp}, 
the decay width $\Gamma_{\SField \rightarrow \pi \pi}$  is taken from Ref.\ \cite{Bugg2007}, 
and the scattering lengths from Ref.\ \cite{PEYAUD200929}.}\label{tab:decay}
\end{table}

The pion-pion scattering parameters in the eLSM have been first calculated in Ref.\ \cite{Parganlija:2010fz}, 
but without a scalar  four-quark state and  without a dynamical scalar glueball. 
 It was found that the pion-pion scattering length 
$m_\pi a_0^0$ is in the range of experimental data only for a small mass of the 
 scalar-isoscalar quarkonium $\sigma$ field, while $m_\pi a_0^2$ agrees well  with
experimental data for  all values of the $\sigma$ mass
[cf.\ Fig.\ 2 in Ref.\ \cite{Parganlija:2010fz}]. Here we examine how the results change if we consider a scalar 
 four-quark state $\chi$ in addition to the quarkonium state $\sigma$.  At first,
we neglect the scalar glueball.

The Lagrangians \eqref{eq:mesonLagrangian} and \eqref{eq:completeTetraquark} contain  ten parameters 
that are of relevance for our fit: $\mu^2, \lambda_1, \lambda_2, c, m_1^2, h_1 + h_2 \equiv h, h_3, 
g_\chiMeson, \gAV, m_\chiMeson$. The parameters $\lambda_2, h_3, c, \mu^2, m_1^2, \gAV$ can be expressed by the 
physical masses of Tab.\ \ref{tab:assignments} or by the remaining model parameters, see 
App.\ \ref{app:massesParameters}. Furthermore, $\lambda_1$ is large-$N_c$ suppressed and is therefore set to zero. 
Thus, only the  three parameters $h, g_\chiMeson, m_\chiMeson$ need to be fitted. 
We used the standard $\chi^2$ procedure to fit the parameters and determine the errors 
($\chi^2 = \chi^2(h, g_\chi, m_\chi)$):
 \begin{align}\label{eq:chi2model1}
    \chi^2 &= \left( \dfrac{\massH - 475 \, \textrm{MeV}}{75  \, \textrm{MeV}} \right)^2 
    + \left( \dfrac{\massS - 1350 \, \textrm{MeV}}{150  \, \textrm{MeV}} \right)^2 \notag \\
    &+ \left( \dfrac{\Gamma_{\HField \rightarrow \pi \pi} - 550 \, \textrm{MeV}}{150  \, \textrm{MeV}} \right)^2 
    + \left( \dfrac{\Gamma_{\SField \rightarrow \pi \pi} -  350  \, 
    \textrm{MeV}}{ 150  \, \textrm{MeV}} \right)^2 \notag \\
    &+ \left( \dfrac{m_\pi a_0^0 - 0.218}{0.02} \right)^2  + \left( \dfrac{m_\pi a_0^2+ 0.046}{0.013} \right)^2 \, .
\end{align}
The result of this fit is presented in Tab.\ \ref{tab:Fit1}.

\begin{table}[ht]
  \centering
  \begin{tabular}[t]{ |c|c| }
  \hline 
  \textbf{Param.} &  \textbf{Value} \\ \hline\hline
  $g_\chiMeson$ & $2.86 \pm 0.53$ MeV\\ \hline
  $h$ & $-0.22 \pm 4.7$ \\ \hline
  $m_\chiMeson$ & $533 \pm 33$ MeV\\ \hline\hline
  $\gAV$ & $-12018 \pm 1365$ MeV \\ \hline
  $\mu^2$ & $-879 \times 10^{3}$ $\textrm{MeV}^2$\\ \hline
  $m_1^2$ & $\approx 775^2$ $\textrm{MeV}^2$\\ \hline
  $c$ & $99 \pm 0.4 \times 10^3 \textrm{MeV}^2$ \\ \hline
  $m_\sigmaMeson$ & $1405$ MeV\\ \hline
  $\chiMeson_0$ & $0.24 \pm 0.02$ MeV\\ \hline
  $\mixingAngle$ & $\approx 0$ \\ \hline
\end{tabular}
\begin{tabular}[t]{ |c|c| }
  \hline 
  \textbf{Observ.} &  \textbf{Value} \\ \hline\hline
  $\massH$ & $533 \pm 33$ MeV\\ \hline
  $\massS$ & $1405$ MeV \\ \hline
  $\Gamma_{H \rightarrow \pi \pi}$ & $504 \pm 148$ MeV \\ \hline
  $\Gamma_{S \rightarrow \pi \pi}$ & $420 \pm 144$ MeV \\ \hline
  $m_\pi a_0^0$ & $0.210 \pm 0.016$ \\ \hline
  $m_\pi a_0^2$ & $-0.027 \pm 0.005$ \\ \hline\hline
  \textbf{$\chi^2$ test} &  \textbf{Value} \\ \hline
  $\chi^2$ & $3.5$ \\ \hline
  $\chi_{\textrm{red}}^2$ & $1.8$  \\ \hline 
\end{tabular}\caption{In the upper left box the fitted parameters are given.  The 
parameters in the 
lower left box are calculated from the fitted parameters. 
 $\theta$ is the mixing angle between $\chi$ and $\sigma$.
In the right box the 
numerical results  for 
the observables are given.}\label{tab:Fit1} 
\end{table}

 Next, we consider the scalar glueball as dynamical field as well. Now, $M_G$ and
$G_0$ are additional fit parameters.
Then the $\chi^2$ function is given as
\begin{align}\label{eq:chi2model2}
    \chi^2 &= \left( \dfrac{\massH - 475 \, \textrm{MeV}}{75 \, \textrm{MeV}} \right)^2 
    + \left( \dfrac{\massS - 1350 \, \textrm{MeV}}{150 \, \textrm{MeV}} \right)^2  \notag \\
    &+ \left( \dfrac{m_{G^\prime} - 1720 \, \textrm{MeV}}{50 \, \textrm{MeV}} \right)^2 
    +\left( \dfrac{\Gamma_{\HField \rightarrow \pi \pi} - 550 \, \textrm{MeV}}{150 \, \textrm{MeV}} \right)^2 \notag\\ 
    &+  \left( \dfrac{\Gamma_{\SField \rightarrow \pi \pi} 
    -  350 \, \textrm{MeV}}{ 150 \, \textrm{MeV}} \right)^2
    + \left( \dfrac{\Gamma_{\GField \rightarrow \pi \pi} - 29.3 \, \textrm{MeV}}{6.5 \, \textrm{MeV}} \right)^2  \notag\\
    &+ \left( \dfrac{m_\pi a_0^0 - 0.218}{0.02} \right)^2  + \left( \dfrac{m_\pi a_0^2 + 0.046}{0.013} \right)^2 \, .
\end{align}
The results of the fit for this scenario are  given in Tab.\ \ref{tab:GlFit1}.

\begin{table}[t]
  \centering
  \begin{tabular}[t]{ |c|c| }
  \hline 
  \textbf{Param.} &  \textbf{Value} \\ \hline\hline
  $g_\chiMeson$ & $3.06 \pm 0.54$ MeV\\ \hline
  $h$ & $5.53 \pm 2.75$ \\ \hline
  $M_G$ & $1564 \pm 84$ MeV \\ \hline
  $G_0$ & $428 \pm 135$ MeV\\ \hline
  $m_\chiMeson$ & $547 \pm 33$ MeV\\ \hline\hline
  $\gAV$ & $-11820 \pm 738$ MeV\\ \hline
  $\mu^2$ & $-873 \times 10^{3}$ $\textrm{MeV}^2$\\ \hline
  $m_1^2$ & $730^2$ $\textrm{MeV}^2$\\ \hline
  $c$ & $99 \times 10^{3} \, \textrm{MeV}^2  $ \\ \hline
  $m_\sigmaMeson$ & $1401 \, \textrm{MeV} $ \\ \hline
  $\chiMeson_0$ & $0.24 \pm 0.02$ MeV\\ \hline
\end{tabular}
  \begin{tabular}[t]{ |c|c| }
  \hline 
  \textbf{Observ.} &  \textbf{Value} \\ \hline\hline
  $\massH$ & $546 \pm 33$ MeV\\ \hline
  $\massS$ & $1238 \pm 113$ MeV \\ \hline
  $\massG$ & $1696 \pm 49$ MeV \\ \hline
  $\Gamma_{H \rightarrow \pi \pi}$ & $539 \pm 148$ MeV \\ \hline
  $\Gamma_{S \rightarrow \pi \pi}$ & $503 \pm 98$ MeV \\ \hline
  $\Gamma_{\GField \rightarrow \pi \pi}$ & $29 \pm 7$ MeV \\ \hline
  $m_\pi a_0^0$ & $0.210 \pm 0.016$ \\ \hline
  $m_\pi a_0^2$ & $-0.028 \pm 0.005$ \\ \hline\hline
  \textbf{$\chi^2$ test} &  \textbf{Value} \\ \hline
  $\chi^2$ & $5.1$ \\ \hline
  $\chi_{\textrm{red}}^2$ & $1.7$  \\ \hline 
\end{tabular}\caption{The result of the fit where the glueball is included.}\label{tab:GlFit1} 
\end{table}

\noindent From this fit the following  scalar-isoscalar mixing matrix is obtained:
\begin{equation}\label{eq:mixingMatrix}
     O^T = \begin{pmatrix}
1.00 & 0.00 & 0.00 \\
0.00 & 0.81 & -0.59 \\
0.00 & 0.59 & 0.81
\end{pmatrix} \, ,
    \end{equation}
which corresponds to the following  admixtures of the physical states:
\begin{align}\label{eq:mixing}
        f_0(500)&: \quad 100\% \, \chiMeson,\quad 0\% \, \sigmaMeson,\quad 0\% \, G \, ,\\
        f_0(1370)&:\quad 0\% \, \chiMeson,\quad 65\% \, \sigmaMeson,\quad 35\% \, G \, ,\\
        f_0(1710)&:\quad 0\% \, \chiMeson,\quad 35\% \, \sigmaMeson,\quad 65\% \, G \, .
\end{align}
Let us briefly discuss these results:
\begin{itemize}
    \item Our aim was to correctly reproduce the masses and decay widths of $f_0(500)$, $f_0(1370)$, and $f_0(1710)$. 
    The fit agrees well with experimental data, only  for the scattering length $m_\pi a_0^2$ 
     the theoretical and experimental error bands overlap just barely (the theoretical value is 
    slightly too large). In both fits, with and without a dynamical glueball, similar parameters are obtained, 
     which results in very similar observables. 
    \item The parameter determining the mixing of the four-quark state with the quarkonium state
     is $-2 g_\chi \varphi/(m_\sigmaMeson^2 - m_\chi^2)$, which is (approximately) zero, 
    explaining why $f_0(500)$ is (almost to) $100\%$ a four-quark state.
    
    \item  Although the value of $g_\chiMeson$ is very small, it is numerically not negligible. 
     In App.\ \ref{app:massesParameters} we show that $\gAV$ is proportional to the 
    inverse of $g_\chiMeson$.
    Thus, $\gAV$ would diverge if we send $g_\chiMeson \rightarrow 0$. 
    \item  Since $g_\chiMeson$ is very small,  
    the coupling $\gAV$ between the scalar 
     four-quark state and the 
    (axial-)vector mesons is  rather large, $\gAV \sim -12$ GeV in both fits.
    \item We also tried to identify $\HField = f_0(980)$, another possible  candidate for a four-quark state, 
    but no reasonable fit results were obtained. Our investigation clearly favors the  (non-strange) 
    scalar  four-quark state to be a light and broad state.
    \item The value of $\Gamma_{\SField \rightarrow \pi \pi}$  used in our fit is somewhat problematic due 
    to its uncertain value in  the literature \citep{Patrignani:2016xqp, Bugg2007}. If we exclude this width from our fit, we obtain  
   $\Gamma_{\SField \rightarrow \pi \pi} \approx 3$ GeV for the  case without dynamical glueball and 
   $\Gamma_{\SField \rightarrow \pi \pi} \approx 750$ MeV for the case with dynamical glueball, while the 
   other observables change only slightly. 
    \item We obtain a fit of similar quality if we assign $\GField = f_0(1500)$, but at a cost of a very large dilaton
    condensate $G_0 > 1.5$ GeV. Thus, we cannot make any prediction about whether $f_0(1500)$ or $f_0(1710)$ is 
     more likely the glueball candidate. This question has been  addressed in a model
    similar to ours \cite{JanowskiGlueball}, where the authors found $f_0(1710)$ to be  the scalar 
    glueball while $f_0(1500)$ was found to be  mostly an $\bar{s}s$ quarkonium state. 
    \item The  elements of the matrix \eqref{eq:mixingMatrix} 
    which correspond to the mixing 
    between $\sigmaMeson$ and $G$  are somewhat larger than those of Ref.\ \cite{JanowskiGlueball}, 
     most likely due to the missing strange scalar-isoscalar $\sigma_{S}$ in our two-flavor model.
    \item We find in both fits very similar values for the pion-pion scattering lengths, indicating that the
     scalar glueball is actually not important for pion-pion scattering, which is not too surprising because of 
      its
     large mass.
    \item We checked that  the pion-pion scattering lengths vanish in the chiral limit, 
    i.e., $m_\pi \rightarrow 0$, as required by low-energy theorems. 
    \item To further underline the importance of a light scalar-isoscalar resonance we can take the limit 
    $g_\chiMeson \rightarrow 0$ and $\gAV \rightarrow 0$ to turn off the  interactions of the 
    four-quark state.
     Performing a fit in this limit leads to $m_\pi a_0^0 = 0.156$ and $m_\pi a_0^2 = -0.044$,  i.e., results
     comparable to those of Ref.\ \cite{Parganlija:2010fz} for large $\sigmaMeson$ masses. 
     This result is obtained for both cases, with and  without a dynamical scalar glueball. 
\end{itemize}

\subsection{Pion-nucleon scattering parameters}\label{sec:baryonResults}

Some of the parameters of the baryon Lagrangian have been already determined in Ref.\ \cite{Gallas} 
and are reported in Tab.\ \ref{tab:NucleonParams}.
 
  \begin{table}[ht]
  \centering
  \begin{tabular}[t]{ |c|c| }
  \hline 
  \textbf{Parameter} &  \textbf{Value} \\ \hline\hline
  $c_1$ & $-3.0 \pm 0.6$  \\ \hline
  $c_2$ & $11.6 \pm 3.6$  \\ \hline
  $Z$ & $1.67 \pm 0.2$ \\ \hline
  $m_0$ & $462 \pm 136$ MeV\\ \hline
\end{tabular}
\caption{Parameters determined by a fit of $g_A^N$, $g_A^{N^*}$, 
$\Gamma_{N^* \rightarrow N \pi}$, $\Gamma_{a_1 \rightarrow \pi \gamma}$, where $N^*$ is assigned to 
$N(1535)$, see Ref.\ \cite{Gallas}.}\label{tab:NucleonParams} 
\end{table}

In Ref.\ \cite{Gallas} the isospin-even and isospin-odd scattering lengths have been calculated
 in a model without a scalar four-quark state and a scalar glueball. The authors found 
 $m_\pi a_0^{(-)} = 0.0834 \pm 0.0087$ for the isospin-odd scattering length, which is 
in surprisingly good agreement with the experimental value, see the first entry of the
last column in Tab.\ \ref{tab:isospin-odd}. However,
there were several errors in Eq.\ (19) of Ref.\ \cite{Gallas}. The correct formula
is given in App.\ \ref{app:piNScatFormulas}, and the correct value is the first entry of the second
column in Tab.\ \ref{tab:isospin-odd}. This value is now outside the experimental error band.

As experimental inputs for the scattering lengths, we use the results of the analysis of Refs.\ 
\cite{BARU2011473, BARU201169}, which are based on experimental results [see e.g.\ \cite{Schroder2001, gotta}] 
and isolate the contributions of isospin-breaking electromagnetic interactions (which are not present in our model).
The remaining scattering parameters are taken from Ref.\ \cite{Matsui}.

 The isospin-even scattering length $m_\pi a_0^{(+)}$ has also been calculated
in Ref.\ \cite{Gallas}. There was also a sign error in Eq.\ (18) of Ref.\ \cite{Gallas}, which changes the
behavior of $m_\pi a_0^{(+)}$ as a function of the parameter $m_1$ as shown in Fig.\ 2 of Ref.\ \cite{Gallas}. 
The correct formula is also given in App.\ \ref{app:piNScatFormulas}. We do not
show the corrected graphs; in the first entry of the second column in
Tab.\ \ref{tab:isospin-even} we simply list 
the correspondig value for $m_1=643$ MeV obtained from the global fit of Ref. \cite{Parganlija}. The theoretical value of the scattering length has the opposite sign compared to the experimental value.

 In addition, here we also calculate isospin-even and isospin-odd scattering volumes and 
effective range parameters  within the set-up of Ref.\ \cite{Gallas}. The corresponding values 
are shown in the second to fourth rows in Tabs.\ \ref{tab:isospin-odd} and \ref{tab:isospin-even}. 
 The isospin-odd scattering volumes and the range parameter deviate by factors of 0.5 to 3 from the
experimental values, the isospin-even scattering volumes by factors of 0.4 to 2, while the isospin-even range parameter
is about a factor 7 too small. Note that 
 all 
scattering parameters have also theoretical errors originating from the  uncertainties of the $\chi^2$ fit in
determining the parameters. However, 
we omitted the errors because they are smaller than 5\% and therefore not a reliable measure of uncertainty.

\begin{table}
  \centering
  \begin{tabular}[t]{ |c|c|c|c| }
  \hline
  \textbf{Parameter} &  \textbf{Value} & \textbf{Experiment} \\ \hline\hline
  $m_\pi \, a_{0}^{(-)}$ & $0.0782$ & $0.0861 \pm 0.0009$\\  \hline
  $m_\pi^3 \,a_{1+}^{(-)}$ & $-0.048$ & $-0.081 \pm 0.002$ \\ \hline
  $m_\pi^3 \,a_{1-}^{(-)}$ & $-0.042$ & $-0.013 \pm 0.003$ \\ \hline
  $m_\pi^3 \,r_{0}^{(-)}$ & $0.022$ & $0.007 \pm 0.005$ \\ \hline 
\end{tabular}
\caption{Isospin-odd scattering parameters.}\label{tab:isospin-odd} 
\end{table}

\begin{table}
  \centering
  \begin{tabular}[t]{ |c|c|c|c| }
  \hline
  \textbf{Parameter} &  \textbf{Value} & \textbf{Experiment} \\ \hline\hline
  $m_\pi \, a_{0}^{(+)}$ & $-0.0083$ & $0.0076 \pm 0.0031$\\  \hline
  $m_\pi^3 \,a_{1+}^{(+)}$ & $0.049$ & $0.130 \pm 0.003$ \\ \hline
  $m_\pi^3 \,a_{1-}^{(+)}$ & $-0.093$ & $-0.056 \pm 0.010$ \\ \hline
  $m_\pi^3 \,r_{0}^{(+)}$ & $0.009$ & $-0.06 \pm 0.02$ \\ \hline 
\end{tabular}
\caption{Isospin-even scattering parameters for $m_1 = 643$ MeV and $m_\sigma = 1370$ MeV
as obtained in the model of Ref.\ \cite{Gallas} (shown are the corrected values).}\label{tab:isospin-even} 
\end{table}

 The theoretical values in Tabs.\ \ref{tab:isospin-odd} and \ref{tab:isospin-even}
were computed without a scalar four-quark state or scalar glueball, i.e., they are just a
correction and extension of the results of Ref.\ \cite{Gallas}. While the isospin-odd scattering parameters
are influenced neither by a scalar four-quark state nor by a scalar glueball, and thus cannot be further improved within our
model, we can still study the question whether the introduction of  these states can at least improve the 
description of the isospin-even scattering parameters.
The explicit calculations of the pion-nucleon scattering amplitudes and the isospin-even scattering parameters are 
 deferred to App.\ \ref{app:piNScatFormulas}. Compared to  the model 
of Ref.\ \cite{Gallas} 
we have three new couplings, $a$, $b$, and $c_N$, the values of which are only constrained by
the linear combination \eqref{eq:m0expression}, which should have the correct value of $m_0$ in
order to reproduce the mass of the nucleon and its chiral partner. In a first step we consider only one parameter 
at a time, while setting the other ones to zero. In this way we distinguish three cases:
\begin{align}
\textrm{A:} & \qquad a = \dfrac{m_0}{\chiMeson_0} = 1897.33 \, ,\; b=\cN=0\, ,\\
\textrm{B:} & \qquad b  = \dfrac{m_0}{G_0} = 1.078 \, , \; a = \cN = 0\, ,\\
\textrm{C:} & \qquad \cN  = 2\dfrac{m_0}{\vev^2} =  \dfrac{0.0388}{\textrm{MeV}} \,, \; a = b = 0\, .
\end{align}

\begin{table}
  \centering
  \begin{tabular}[t]{ |c|c|c| }
  \hline
  \textbf{Parameter} &  \textbf{Value} & \textbf{Experiment} \\ \hline\hline
  $m_\pi \, a_{0}^{(+)}$& $11.196$ & $0.0076 \pm 0.0031$ \\  \hline
  $m_\pi^3 \,a_{1+}^{(+)}$ & $-3.422$ & $0.133 \pm 0.004$ \\ \hline
  $m_\pi^3 \,a_{1-}^{(+)}$ & $-3.365$ & $-0.056 \pm 0.010$ \\ \hline
  $m_\pi^3 \,r_{0}^{(+)}$  & $9.061$ & $-0.06 \pm 0.02$ \\ \hline 
\end{tabular}
\caption{Results with $a = 1897.33$, $b = \cN = 0$.}\label{tab:baryonFit1} 
\end{table}

\begin{table}
  \centering
  \begin{tabular}[t]{ |c|c|c| }
  \hline
  \textbf{Parameter}  &  \textbf{Value} & \textbf{Experiment} \\ \hline\hline
  $m_\pi \, a_{0}^{(+)}$& $-0.0079$ & $0.0076 \pm 0.0031$ \\  \hline
  $m_\pi^3 \,a_{1+}^{(+)}$  & $0.048$ & $0.133 \pm 0.004$ \\ \hline
  $m_\pi^3 \,a_{1-}^{(+)}$& $-0.091$ & $-0.056 \pm 0.010$ \\ \hline
  $m_\pi^3 \,r_{0}^{(+)}$  & $0.009$ & $-0.06 \pm 0.02$ \\ \hline 
\end{tabular}
\caption{Results with $b = 1.078$, $a = \cN = 0$.}\label{tab:baryonFit2} 
\end{table}

\begin{table}
  \centering
  \begin{tabular}[t]{ |c|c|c| }
  \hline
  \textbf{Parameter}  &  \textbf{Value} & \textbf{Experiment} \\ \hline\hline
  $m_\pi \, a_{0}^{(+)}$& $-0.0078$ & $0.0076 \pm 0.0031$ \\  \hline
  $m_\pi^3 \,a_{1+}^{(+)}$  & $0.048$ & $0.133 \pm 0.004$ \\ \hline
  $m_\pi^3 \,a_{1-}^{(+)}$& $-0.091$ & $-0.056 \pm 0.010$ \\ \hline
  $m_\pi^3 \,r_{0}^{(+)}$  & $0.008$ & $-0.06 \pm 0.02$ \\ \hline 
\end{tabular}
\caption{Results with $\cN = \dfrac{0.0388}{\textrm{MeV}}$, $a = b = 0$.}\label{tab:baryonFit3} 
\end{table}

In Tab.\ \ref{tab:baryonFit1} we consider case A, where only the scalar four-quark state contributes to the 
explicit mass term.  In this case the scalar glueball contributes only indirectly to the 
pion-nucleon scattering parameters via the $G-\sigmaMeson$ and $G-\chiMeson$ mixing. Due to the small 
value of the four-quark
condensate found in the fit of the meson sector the coupling between the scalar four-quark
state and the nucleons must be extremely large in order to obtain $m_0 = 462$ MeV. 
This leads to scattering parameters that are off by several orders of magnitude. 

In Tabs.\ \ref{tab:baryonFit2} and \ref{tab:baryonFit3} we consider the cases B and C, where $m_0$ originates  
 either exclusively from the gluon condensate or from the anomaly contribution, respectively. 
 In these cases, the scalar four-quark state contributes to pion-nucleon scattering only via 
the $\chiMeson-\sigmaMeson$ and $\chiMeson-G$ coupling.
The results  are rather similar for both cases, and the numerical values rather close
to those of the model of Ref.\ \cite{Gallas}, cf.\ Tab.\ \ref{tab:isospin-even}. This is expected because 
 on the one hand the glueball is rather heavy and thus cannot substantially influence the 
scattering parameters 
in case B. On the other hand, the additional anomalous contribution $\sim c_N \varphi \simeq 6$
to the $\sigma N N^*$ coupling is about a factor of 2 smaller in magnitude than the Yukawa coupling
$(\hat{g}_{1} \Exp{\delta} - \hat{g}_{2} \Exp{-\delta})/4$, and thus case C is, as far as pion-nucleon scattering is concerned, essentially identical to the model of
Ref.\ \cite{Gallas}. Case A is markedly different from cases B and C because of the presence of
the light four-quark field $f_0(500)$, which has a substantial impact on pion-nucleon scattering parameters.

We now perform a simultaneous $\chi^2$ fit for all three parameters $a$, $b$, and $c_N$,
respecting the constraint \eqref{eq:m0expression}. Using the constraint that $b$ is positive yields the results of 
Tab.\ \ref{tab:baryonFit4}. 
It should also be noted that  we find basically the same results if we set either $b = 0$ or $\cN = 0$.
It is important, however, that $a$ is non-vanishing. Although the agreement between the theoretically calculated 
scattering parameters
and experimental data is now in general better, the values still deviate by factors of 1.5 to 3 
(and the scattering length $a_0^{(+)}$ and the range parameter have the wrong sign).

\begin{table}
  \centering
  \begin{tabular}[t]{ |c|c|c|c| }
  \hline
  \textbf{Parameter}  &  \textbf{Value} & \textbf{Experiment} \\ \hline\hline
  $m_\pi \, a_{0}^{(+)}$ & $-0.0029$ & $0.0076 \pm 0.0031$ \\  \hline
  $m_\pi^3 \,a_{1+}^{(+)}$  & $0.047$ & $0.133 \pm 0.004$ \\ \hline
  $m_\pi^3 \,a_{1-}^{(+)}$  & $-0.092$ & $-0.056 \pm 0.010$ \\ \hline
  $m_\pi^3 \,r_{0}^{(+)}$ & $0.012$ & $-0.06 \pm 0.02$ \\ \hline 
\end{tabular}
\caption{Best fit where $a = 1.237$, $b = 1.036$, $\cN = 0.0015$ 
(the glueball contribution dominates).}\label{tab:baryonFit4} 
\end{table}

\section{Summary and Discussion}
\label{sec:discussion}

In this paper, we studied the influence of the light four-quark state $f_{0}(500)$ 
and the scalar glueball on pion-pion and pion-nucleon scattering in the framework of the
eLSM for $N_\textrm{f} = 2$ flavors. We first performed a $\chi^{2}$ fit to properties in the mesonic sector.
We found a physically acceptable minimum for which the resonance
$f_{0}(500)$ is  (almost) exclusively a four-quark state, $f_{0}(1370)$ 
 predominantly a light
quark-antiquark state, and $f_{0}(1710)$ predominantly a gluonic state. The masses and
the decay widths of these resonances as well as pion-pion scattering lengths can be
correctly described: in particular $a_{0}^{0}$ is strongly dependent on the
additional attraction  on account of $f_{0}(500)$, and the
presence of the latter is necessary for a good description of the data. The
resonance $f_{0}(1710)$ is quite heavy, hence it does not affect the
results in a substantial way, but its presence is nevertheless important to stabilize the fit
(smaller errors, reasonable upper limit for the gluon condensate $G_0$). Another
notable result is the observation that the (axial-)vectors turn out to
interact quite strongly with the scalar four-quark state ($\gAV/G_{0}\sim
-30$), while the coupling to the (pseudo)scalar quarkonia is rather small
($g_{\chi}/G_{0}\sim 0$).

 Then we studied the role of $f_{0}(500)$ and $f_0(1710)$ in the
baryonic sector of the eLSM.  The nucleon and its chiral
partner were incorporated into the model in the so-called mirror assignment and theoretical expressions for the
pion-nucleon scattering parameters, namely the isospin-even and isospin-odd
scattering lengths, scattering volumes, and effective range parameters were derived. First, we
presented results without the scalar  four-quark and glueball state,
correcting and extending results of Ref.\ \cite{Gallas}. 
Our results are found to be in the correct order of magnitude compared to 
experimental data but there is still room for improvement. 

For  instance,  the  inclusion  of  the  $\Delta$ resonance  is  expected  to  be important, since this state is just slightly 
heavier than the nucleon and it couples strongly to $N\pi$. Indeed, as a preliminary study shows \cite{schneider}, the 
effect of the $\Delta$ pushes $a_0^{(+)}$ toward positive values,  in agreement with Refs.\ 
\citep{BARU2011473, BARU201169}. A proper study of this issue would require the inclusion of the $\Delta$ and its 
chiral partner into the eLSM in the framework of the mirror assignment.

Another straightforward extension of this work would be to consider the three-flavor
case. Two additional resonances appear in the scalar-isoscalar sector: the
strange-antistrange quarkonium (predominantly $f_{0}(1500))$ and the four-quark
state $f_{0}(980)$ (a kaon-kaon state in the molecular picture, a $[u,s][\bar{u},\bar{s}]+[d,s][\bar{d},\bar{s}]$ state
in the tetraquark picture). Moreover, also the quarkonium states $a_{0}(1450)$ and $K_{0}^{\ast}(1430)$
and the four-quark states $a_{0}(980)$ and $K_{0}^{\ast}(800)$ would enter
this extended scenario. In this context, the difference between
different internal structures of the four-quark states (meson-meson or diquark-antidiquark) would also
become visible in terms of different Clebsch-Gordon coefficients.

In the baryonic sector, an interesting future work would be to continue the
nonzero-density study of Ref.\ \cite{tqfinitedensity}. Due to the recent
discovery of gravitational waves emitted by neutron-star binary mergers, it is expected that
the equation of state of nuclear matter at high
density can be more precisely determined in the future.

\begin{acknowledgments}
The authors thank the \textit{Chiral Field Theory Group}, P. Kovacs, Daiki Suenaga, and S. Rechenberger for useful discussions.
F.G.\ acknowledges financial support from the Polish
National Science Centre NCN through the OPUS project no.\ 2015/17/B/ST2/01625.
F.G.\ and D.H.R.\ acknowledge financial support from DFG under grant no.\ RI 1 1181/6-1.
D.H.R.\ is partially supported by the High-end Foreign 
Experts project GDW20167100136 of the State
Administration of Foreign Experts Affairs of China.
\end{acknowledgments}

\section{Appendices}
\label{sec:appendix}

We used \textit{Mathematica}\textsuperscript{\textregistered} for the numerical evaluation and the fits 
performed in this work.
The notebooks can be found on github: \href{https://github.com/Phillip2/eLSM}{https://github.com/Phillip2/eLSM}

\appendix

\section{Tetraquark-baryon interaction terms}
\label{app:tetraquarkBaryon} 

As in the meson sector, we construct possible tetraquark-baryon interaction terms for $N_\textrm{f} = 3$ and 
then reduce them to $N_\textrm{f} = 2$ in order to verify that the effective interaction of 
Eq.\ \eqref{eq:gallasTetraquarkNucleon} can be derived within our approach.
To this end, we use the formalism of Ref.\ \cite{Olbrich:2015gln}, where four baryonic multiplets, 
$N_1, N_2, M_1$, and $M_2$ were constructed using the quark-diquark picture for baryons,  cf.\ Tab.\ III 
of Ref.\ \cite{Olbrich:2017fsd} for the transformation properties under chiral symmetry, parity, and charge conjugation.
We find the following terms  where scalar tetraquarks couple to baryons:
\begin{align}
\mathcal{L}_{\chiMeson B} & = \kappa_1 \Trace{\bar{M}_{1\trm{R}} D_{\trm{L}}D_{\trm{L}}^\dag N_{1\trm{L}} 
+ \bar{M}_{2\trm{L}} D_{\trm{R}}D_{\trm{R}}^\dag N_{2\trm{R}} + \hc} \notag \\
& + \kappa_2 \Trace{\bar{M}_{1\trm{L}} D_{\trm{R}}D_{\trm{R}}^\dag N_{1\trm{R}} 
+ \bar{M}_{2\trm{R}} D_{\trm{L}}D_{\trm{L}}^\dag N_{2\trm{L}} + \hc} \notag \\
&+ \left\{ \kappa_3  \Trace{\bar{M}_{1\trm{R}} N_{1\trm{L}} + \bar{M}_{2\trm{L}} N_{2\trm{R}} + \hc} \right.  \notag \\
&\hspace*{0.15cm} +\left. \kappa_4 \Trace{\bar{M}_{1\trm{L}} N_{1\trm{R}} 
+ \bar{M}_{2\trm{R}} N_{2\trm{L}} + \hc} \right\} \notag\\ 
& \times \Trace{D_{\trm{R}}D_{\trm{R}}^\dag+D_{\trm{L}}D_{\trm{L}}^\dag}\, ,
\end{align}
which reduces to Eq.\ \eqref{eq:gallasTetraquarkNucleon} for two flavors using the same approach as in Sec.\
\ref{sec:tetraquarks}. We show this  explicitly for one term:
\begin{align}
    & \Trace{\bar{M}_{2\trm{L}} D_{\trm{R}}D_{\trm{R}}^\dag N_{2\trm{R}}} 
    = \Trace{\bar{M}_{2} D_{\trm{R}}D_{\trm{R}}^\dag N_{2}} + \ldots\notag \\
    = \, &\dfrac{1}{2} \Trace{(\bar{B}_{M^*} - \bar{B}_M) D_{\trm{R}}D_{\trm{R}}^\dag (B_{N^*} - B_{N})} + \ldots \notag \\
    = \,  & \dfrac{1}{4} \Trace{(\bar{B}_{M^*} - \bar{B}_M) (D D^\dag + \ldots ) (B_{N^*} - B_{N})} + \ldots \notag \\
    = \, & \dfrac{1}{4} T_{ab} \Trace{ \bar{B}_{M^*}  A_b A_a^T  B_N} + \ldots \notag \\
   \stackrel{N_\textrm{f} = 2}{\longrightarrow} \, &\dfrac{1}{4}\, \chiMeson\, \bar{\psi}_2 \psi_1+ \ldots \, .
\end{align}
From the first to the second line we used the definition of the baryon fields with definite 
parity and charge-conjugation properties as 
defined in Ref.\ \cite{Olbrich:2017fsd}. The $N_\textrm{f} = 2$ limit in the last line is obtained by setting $a = b = 3$ and by 
reducing
\begin{equation}
    B_{M^*} \stackrel{ N_\textrm{f}=2}{\longrightarrow} \begin{pmatrix}
0 & 0 & \psi_{2,1} \\
0 & 0 & \psi_{2,2} \\
0 & 0 & 0
\end{pmatrix} \, , \quad
B_{N} \stackrel{ N_\textrm{f}=2}{\longrightarrow} \begin{pmatrix}
0 & 0 & \psi_{1,1} \\
0 & 0 & \psi_{1,2} \\
0 & 0 & 0
\end{pmatrix} \, .
    \end{equation}

\section{Masses and Parameters}
\label{app:massesParameters}

\subsection{Axial-vector--pseudoscalar mixing}

After spontaneous symmetry breaking, the  field $\sigmaMeson$ is shifted by  its 
vacuum expectation value, i.e., $\sigmaMeson \rightarrow \vev + \sigmaMeson$. 
This yields mixing terms between axial-vector and 
pseudoscalar mesons,  e.g.\ $\sim \vec{a}_1^\mu \cdot \partial_\mu \vec{\pi}$. 
Such terms can be eliminated by a redefinition of the fields,
\begin{align}
    \vec{a}_1^\mu &\rightarrow \vec{a}_1^\mu + Z w \,\partial^\mu \vec{\pi} \, , \notag \\
    \vec{\pi} & \rightarrow Z \vec{\pi} \, ,
\end{align}
 where
\begin{equation}
w =  \dfrac{g_1 \vev}{m_{a_1}^2} \, ,
\end{equation}
with $m_{a_1}^2$ from Eq.\ (\ref{eq:a1mass}), and
\begin{equation}
    Z =  \left(1 - \dfrac{g_1^2 \vev^2}{ m_{a_1}^2}\right)^{-1/2}.
\end{equation}

\subsection{Masses}\label{app:masses}

The masses of the resonances are calculated from the second partial derivative of the potential density 
(which is obtained from Eqs.\ \eqref{eq:mesonLagrangian} and \eqref{eq:completeTetraquark}) with respect to 
the corresponding fields:
\begin{align}
 m_{\sigma}^{2} & =  \mu^{2} - c + 3 \left( \lambda_{1} 
+ \dfrac{\lambda_{2}}{2} \right) \varphi^{2} -2 g_{\chiMeson} \chiMeson_0 \, ,\label{eq:sigmamass} \\
 m_{\pi}^{2} & = Z^{2} \left[ \mu^{2} - c + \left( \lambda_{1} 
+ \dfrac{\lambda_{2}}{2} \right) \varphi^{2} -2 g_{\chiMeson} \chiMeson_0\right]  \, , \label{eq:pimass} \\
 m_{\eta}^{2} & = Z^{2} \left[ \mu^{2} + c + \left( \lambda_{1} 
+ \dfrac{\lambda_{2}}{2} \right) \varphi^{2} + 2 g_{\chiMeson} \chiMeson_0 \right] \, , \label{eq:etamass} \\
 m_{a_{0}}^{2} & = \mu^{2} + c + \left( \lambda_{1} 
+ \dfrac{3 \lambda_{2}}{2} \right) \varphi^{2} +2 g_{\chiMeson} \chiMeson_0 \, , \label{eq:a0mass} \\
 m_{\rho}^{2} &  = m_{1}^{2} 
+  ( h + h_{3})\dfrac{\varphi^{2}}{2}+2 \gAV \chiMeson_0  \, ,
\label{eq:rhomass} \\
 m_{\omega}^2 & =m_{1}^{2} 
+  ( h + h_{3}) \dfrac{\varphi^{2}}{2}  \, -\,  2 \gAV \chiMeson_0 \, \label{eq:omegamass}\\
 m_{a_{1}}^{2} & 
=  m_{1}^{2} + ( 2 g_1^2 +  h - h_{3})\dfrac{\varphi^{2}}{2} 
+2 \gAV \chiMeson_0  \,, \label{eq:a1mass} \\
 m_{f_{1}}^{2} & =  m_{1}^{2} 
+  ( 2 g_1^2 +  h - h_{3})\dfrac{\varphi^{2}}{2}
  \, -\,  2 \gAV \chiMeson_0 \, . \label{eq:f1mass}
\end{align}

\subsection{Parameters}\label{app:parameters}

The  vacuum expectation value of $\sigmaMeson$, $\vev = Z f_\pi$, is determined from the axial current. 
The  vacuum expectation value
of the scalar  four-quark state is determined by the minimum of the potential density:
\begin{equation} \label{B18}
    \chiMeson_0(g_\chiMeson, m_\chiMeson^2) = \dfrac{g_\chi \vev^2}{m_{\chiMeson}^2} \, .
\end{equation}
We can use the masses, whose values are given by experimental data, to fix some of the model parameters
\begin{align}
\lambda_2 &= \dfrac{1}{\vev^2} \left( m_{a_0}^2 - \dfrac{m_\etaMeson^2 }{Z^2} \right)\, , \label{eq:param1} \\
	    h_3 &= \dfrac{1}{\vev^2} \left( m_\rho^2 - m_{a_1}^2 + g_1^2 \vev^2 \right) \, ,\\
    c &= c(g_\chiMeson, m_\chiMeson^2) = \dfrac{1}{2 Z^2} \left(m_\etaMeson^2 - m_\pi^2 \right) 
    - 2 g_\chiMeson \chiMeson_0(g_\chiMeson, m_\chiMeson^2) \, , \\
    \mu^2 &= \mu^2(\lambda_1) = \dfrac{1}{2} \left[ \dfrac{1}{Z^2} (m_\etaMeson^2 + m_\pi^2) 
    - \vev^2 (2 \lambda_1 + \lambda_2) \right]  \, , \\
    m_1^2 &= m_1^2(g_\chiMeson, m_\chiMeson^2, h) \notag \\
    = \dfrac{1}{2} &\left[ m_\rho^2 + m_{a_1}^2 - 4 \gAV(g_\chiMeson, m_\chiMeson^2)  
    \chiMeson_0(g_\chiMeson, m_\chiMeson^2) 
    - \vev^2 (g_1^2 +  h) \right] \, ,\label{eq:param5} \\
     \gAV &= \gAV( g_\chiMeson, m_\chiMeson^2) = \dfrac{m_\rho^2 - m_\omega^2}{
     4 \chiMeson_0(g_\chiMeson, m_\chiMeson^2)} = m_\chiMeson^2 
     \dfrac{m_\rho^2 - m_\omega^2}{4 g_\chiMeson \vev^2} \, ,
\end{align}  
where we used Eq.\ (\ref{B18}) in the last step of the last line. Note that $\gAV$ and $g_\chiMeson$ are
inversely proportional to each other, i.e., a small value of $g_\chiMeson$ requires a large value of $\gAV$ and vice versa.

\section{Meson sector  without dynamical scalar glueball}
\label{app:meson}

\subsection{ Scalar mixing angle}

The shift $\sigmaMeson \rightarrow \vev + \sigmaMeson$ and $\chiMeson \rightarrow \chiMeson_0 + \chiMeson$ 
leads to a non-diagonal mass matrix. We  rotate the fields via an $SO(2)$ transformation
\begin{equation}\label{eq:rotation}
\begin{pmatrix}
\chiMeson \\
\sigmaMeson
\end{pmatrix}
 = \begin{pmatrix}
\cos \mixingAngle & -\sin \mixingAngle  \\
\sin \mixingAngle & \cos \mixingAngle 
\end{pmatrix} 
\begin{pmatrix}
H \\
S
\end{pmatrix} \, ,
\end{equation}
and demand that the mass matrix in the basis of the new fields, $\HField$ and $\SField$, must be diagonal. 
This leads to  a mixing angle
\begin{equation}
    \mixingAngle = \dfrac{1}{2} \arctan \dfrac{4 g_{\chiMeson} \vev}{ m_{\sigmaMeson}^2 
    - m_{\chiMeson}^2} \, .
\end{equation}
We then relate the physical masses to the unphysical ones:
\begin{align}\label{eq:physicalMasses}
 M_H^2 = m_{\chiMeson}^2 \cos^2 \mixingAngle + m_{\sigmaMeson}^2 
\sin^2 \mixingAngle - 2 g_{\chiMeson} \vev \sin 2\mixingAngle \, , \\
 M_S^2 =  m_{\sigmaMeson}^2 \cos^2 \mixingAngle 
+m_{\chiMeson}^2 \sin^2 \mixingAngle + 2 g_{\chiMeson} \vev \sin 2\mixingAngle \, . \label{eq:physicalMassesEnd}
\end{align}

\subsection{Decay widths}

We are interested in the decay of the scalars $\sigmaMeson$ and $\chiMeson$ into two pions. 
The information of this decay is contained in the Lagrangians
\begin{align}
\mathcal{L}_{\sigmaMeson \rightarrow \pi \pi} &= A_\sigmaMeson \sigma \vec{\pi}^{2} 
+ B_\sigmaMeson \sigma \partial_{\mu} \vec{\pi} \cdot \partial^{\mu} \vec{\pi} 
- C_\sigmaMeson \, \sigmaMeson \vec{\pi} \cdot \Box \vec{\pi} \, , \\
\mathcal{L}_{\chiMeson \rightarrow \pi \pi} &= A_\chiMeson \chiMeson \vec{\pi}^{2} 
+ B_\chiMeson \chiMeson \partial_{\mu} \vec{\pi} \cdot \partial^{\mu} \vec{\pi}\, ,
\end{align}
where
\begin{align}
A_{\sigmaMeson} &= -  Z^2 \vev \left( \lambda_1 + \dfrac{\lambda_2}{2} \right) \, ,  \label{eq:Asigma}\\
B_{\sigmaMeson} &= Z^2 w \left[ -2 g_1  + \dfrac{w \vev}{2} (2 g_1^2  + h - h_3 ) \right] \, , 
 \label{eq:Bsigma} \\
C_{\sigmaMeson} &= -g_1 Z^2 w \, , \label{eq:Csigma} \\
A_{\chiMeson} &= g_\chiMeson Z^2 \, , \\
B_{\chiMeson} &= \gAV w^2 Z^2 \, .
\end{align}
The Feynman amplitudes of the physical fields are obtained from the mixing:
\begin{align}
    \mathcal{M}_{\HField \pi \pi} &= \mathcal{M}_{\sigmaMeson \pi \pi}(m_H) \sin \mixingAngle 
    + \mathcal{M}_{\chiMeson \pi \pi}(m_H) \cos \mixingAngle \, , \\
    \mathcal{M}_{\SField \pi \pi} &= \mathcal{M}_{\sigmaMeson \pi \pi}(m_S) \cos \mixingAngle 
    - \mathcal{M}_{\chiMeson \pi \pi}(m_S) \sin \mixingAngle \, .
\end{align}
where
\begin{align}
-i \mathcal{M}_{\sigmaMeson \pi \pi}(m_X) &= i \left( A_{\sigmaMeson} 
- B_{\sigmaMeson} \dfrac{ m_{X}^2 - 2 m_{\pi}^2}{2}  - C_{\sigmaMeson} m_{\pi}^2 \right) \, , \\
-i \mathcal{M}_{\chiMeson \pi \pi}(m_X) &= i \left( A_{\chiMeson} 
- B_{\chiMeson} \dfrac{ m_{X}^2 - 2 m_{\pi}^2}{2} \right)  \, .
\end{align}
The decay widths are given by:
\begin{align}\label{eq:physicalDecayWidths}
\Gamma_{\HField\rightarrow \pi\pi} &= 3 \dfrac{k_f(m_\HField,m_\pi,m_\pi)}{4 \pi m_{\HField}^2} 
\left| -i \mathcal{M}_{\HField \pi \pi} \right|^2 \, , \\
\Gamma_{\SField\rightarrow \pi\pi} &= 3 \dfrac{k_f(m_\SField,m_\pi,m_\pi)}{4 \pi m_{\SField}^2} 
\left| -i \mathcal{M}_{\SField \pi \pi} \right|^2 \, . 
\end{align}

\subsection{Pion-pion scattering}

\begin{figure}[ht] 

\includegraphics{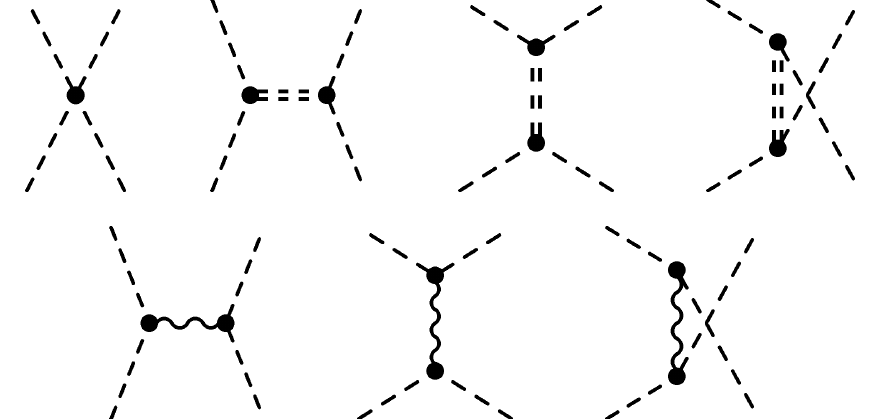}

\caption{The dashed lines correspond to the pion, the wavy line to the $\rho$ meson, and the double-dashed line to 
the $\sigmaMeson$ or the $\chiMeson$. (Each diagram with internal double-dashed lines occurs twice, one with 
the exchange of $\sigmaMeson$ and one with the exchange of $\chiMeson$.)}
\label{fig:treeLevelDiagramsPionPion}
\end{figure}

The scattering amplitude is calculated from the tree-level amplitudes in Fig.\ \ref{fig:treeLevelDiagramsPionPion}.
The pion-pion interaction consists of three parts,
\begin{equation}
\mathcal{L}_{\pi \pi} = \mathcal{L}_{4\pi} + \mathcal{L}_{\sigmaMeson \pi \pi} + \mathcal{L}_{\rho \pi \pi} \, ,
\end{equation}
which can be extracted from the eLSM Lagrangian \eqref{eq:mesonLagrangian}.
From this we obtain the scattering amplitude 
\begin{align}
\mathcal{M}_{\pi \pi} (s,t,u) &= i \delta_{ab} \delta_{cd} A(s,t,u) + i \delta_{ac} \delta_{bd} A(t,u,s) \notag \\
&+ i \delta_{ad} \delta_{bc} A(u,s,t) \, ,
\end{align}
where
\begin{eqnarray}
\lefteqn{A(s,t,u) = (g_1^2 - h_3) Z^4 w^2 s - 2 \left( \lambda_1 + \dfrac{\lambda_2}{2} \right) Z^4 }\notag \\
&- &( h + h_3) Z^4 w^2 (s - 2 m_\pi^2) \notag \\
&- &\dfrac{1}{s - m_{\sigmaMeson}^2} \left[ - 2 m_\pi^2 C_{\sigmaMeson  } + B_{\sigmaMeson  } (2 m_\pi^2 - s) 
+ 2 A_{\sigmaMeson  }  \right]^2  \notag \\
&+& \left( A_{\rho  } + B_{\rho  } \dfrac{t}{2} \right)^2 \dfrac{u - s}{t - m_\rho^2} + \left(  A_{\rho  } 
+ B_{\rho  } \dfrac{u}{2} \right)^2 \dfrac{t - s}{u - m_{\rho}^2} \, .\notag \\ \label{eq:C20}
\end{eqnarray}
The coefficients $A_{\sigmaMeson }$, $B_{\sigmaMeson }$, and $C_{\sigmaMeson }$ are  given in 
Eqs.\ \eqref{eq:Asigma}--\eqref{eq:Csigma},  while
$A_\rho = g_{1} Z^{2} m_{\rho}^{2}/m_{a_{1}}^{2}$  and $B_\rho = g_{2} Z^{2} w^{2}$.

The scattering amplitude in the  $I = 0$ channel is given by the relation \cite{Gasser}
\begin{equation}
 T^0 (s,t,u)= 3 A(s,t,u)+ A(t,u,s)+ A(u,s,t) \, ,
\end{equation}
 from which the isospin-zero scattering length is  computed as 
\begin{equation}
m_\pi a_0^0 = \dfrac{1}{32 \pi} T^0  (4 m_\pi^2, 0,0) \, .
\end{equation}

On the other hand, the scattering amplitude for isospin $I = 2$ is given by
\begin{equation}
 T^2 (s,t,u) = 2 A(u,s,t) \, ,
\end{equation}
and the $s$-wave scattering length is extracted as
\begin{equation}
m_\pi a_0^2 = \dfrac{1}{32 \pi} T^2  (4 m_\pi^2,0,0) \, .
\end{equation}

After the introduction of the scalar  four-quark field, the term
\begin{equation}
\dfrac{1}{s - m_{\sigmaMeson}^2} \left[ - 2 m_\pi^2 C_{\sigmaMeson  } + B_{\sigmaMeson  } (2 m_\pi^2 - s) 
+ 2 A_{\sigmaMeson  }  \right]^2 \, ,
\end{equation}
in Eq.\ \eqref{eq:C20} is replaced by
\begin{align}\label{eq:mHmS}
& \dfrac{1}{s -  M_{\HField}^2} \left[ - 2 m_\pi^2 C_{\HField  } + B_{\HField  } (2 m_\pi^2 - s) 
+ 2 A_{\HField  }  \right]^2 \notag\\ 
+ & \dfrac{1}{s -  M_{\SField}^2} \left[ - 2 m_\pi^2 C_{\SField  } + B_{\SField  } (2 m_\pi^2 - s) 
+ 2 A_{\SField  }  \right]^2 \, .
\end{align}
The new coefficients are given as
\begin{align}
A_{\HField  } &= A_{\sigmaMeson  } \sin \, \mixingAngle +  A_{\chiMeson  } \cos \, \mixingAngle \, , \\
A_{\SField  } &= A_{\sigmaMeson  } \cos \, \mixingAngle -  A_{\chiMeson  } \sin \, \mixingAngle \, , \\
B_{\HField  } &= B_{\sigmaMeson  } \sin \, \mixingAngle +  B_{\chiMeson  } \cos \, \mixingAngle  \, ,\\
B_{\SField  } &= B_{\sigmaMeson  } \cos \, \mixingAngle -  B_{\chiMeson  } \sin \, \mixingAngle \, , \\
C_{\HField  } &= C_{\sigmaMeson  } \sin \, \mixingAngle \, , \\
C_{\SField  } &= C_{\sigmaMeson  } \cos \, \mixingAngle \, .
\end{align}
Thus, at threshold ($s \equiv 4 m_{\pi}^2, \, t = 0, \,  u = 0$), we obtain the scattering lengths
\begin{widetext}

\begin{align}
m_\pi a_0^0 &= \dfrac{1}{32 \pi} \Bigg \lbrace 12 (g_1^2 - h_3) Z^4 w^2 m_{\pi}^2 
- 10 \left( \lambda_1 + \dfrac{\lambda_2}{2} \right) Z^4 - 2 ( h + h_3) Z^4 w^2 m_{\pi}^2 \notag \\
&+ \dfrac{12}{ \massH^2 - 4 m_\pi^2} \left[ (B_{\HField  } + C_{\HField  } ) m_{\pi}^2 - A_{\HField  } \right]^2 
+ \dfrac{12}{ \massS^2 - 4 m_\pi^2} \left[ (B_{\SField  } + C_{\SField  } ) m_{\pi}^2 
- A_{\SField  } \right]^2 \notag\\
&+ \dfrac{8}{ \massH^2} \left[ (B_{\HField  } - C_{\HField  } ) m_{\pi}^2 + A_{\HField  } \right]^2 
+ \dfrac{8}{ \massS^2} \left[ (B_{\SField  } - C_{\SField  } ) m_{\pi}^2 + A_{\SField  } \right]^2 
+ 16 g_1^2 Z^4 m_{\pi}^2 \dfrac{m_\rho^2}{m_{a_1}^4}\Bigg \rbrace \, , \label{eq:a00} 
\end{align}
and
\begin{align}
m_\pi a_0^2 &= \dfrac{1}{16 \pi} \Bigg \lbrace -2 \left( \lambda_1 + \dfrac{\lambda_2}{2} \right) Z^4 
+ 2 (h_1 + h_2 + h_3) Z^4 w^2 m_{\pi}^2 - 4 g_1^2 Z^4 m_\pi^2 \dfrac{m_\rho^2}{m_{a_1}^4} \notag \\
&+ \dfrac{4}{ \massH^2} \left[ (B_{\HField  } - C_{\HField  } ) m_{\pi}^2 + A_{\HField  } \right]^2 
+ \dfrac{4}{ \massS^2} \left[ (B_{\SField  } - C_{\SField  } ) m_{\pi}^2 + A_{\SField  } \right]^2\Bigg 
\rbrace \, .\label{eq:a02}
\end{align}

\end{widetext}

\section{Meson sector with  dynamical scalar glueball}
\label{app:glueball}

\subsection{Decay widths}

From Eqs.\ \eqref{eq:mesonLagrangian} and \eqref{eq:completeTetraquark} we read off the relevant terms for the 
decay $\Gamma_{G \rightarrow \pi \pi}$:
\begin{align}
\mathcal{L}_{\textrm{Glueball-Int}} &= - \dfrac{Z^2}{G_0} \left(  \mu^2 
-  g_\chiMeson \chiMeson_0 \right) G \vec{\pi}^2 \notag \\
&+  \dfrac{Z^2 w^2}{G_0} \left( m_1^2 + \gAV \chiMeson_0 \right) G (\partial_\mu \pi)^2 \, .
\end{align}
This yields another contribution which is of the same form as the $\sigmaMeson \vec{\pi}^2$ and the 
$\chiMeson \vec{\pi}^2$ interaction. The corresponding partial amplitudes are
\begin{align}
A_{G  } &= - \dfrac{Z^2}{G_0} \left(\mu^2 - g_\chiMeson \chiMeson_0\right) \, , \notag \\
B_{G  } &= \dfrac{Z^2 w^2}{G_0} \left( m_1^2 + \gAV \chiMeson_0 \right) \, .
\end{align}
We define the inverse mixing matrix as  $Q = O^T$ (see Eq.\ \eqref{eq:threeRotation}) 
such that the new coefficients are given as 
\begingroup
\allowdisplaybreaks
\begin{align}
A_{\HField  } &= \QQ{11} \, A_{\chiMeson  }  + \QQ{21} \, A_{\sigmaMeson  }  + \QQ{31} \, A_{G  }\, , \\
A_{\SField  } &=  \QQ{12} \, A_{\chiMeson  }  + \QQ{22} \, A_{\sigmaMeson  }  + \QQ{32} \, A_{G  }\, , \\
A_{\GField  } &=  \QQ{13} \, A_{\chiMeson  }  + \QQ{23} \, A_{\sigmaMeson  }  + \QQ{33} \, A_{G  }\, , \\
B_{\HField  } &=  \QQ{11} \, B_{\chiMeson  }  + \QQ{21} \, B_{\sigmaMeson  }  + \QQ{31} \, B_{G  }\, , \\
B_{\SField  } &= \QQ{12} \, B_{\chiMeson  }  + \QQ{22} \, B_{\sigmaMeson  }  + \QQ{32} \, B_{G  }\, , \\
B_{\GField  } &= \QQ{13} \, B_{\chiMeson  }  + \QQ{23} \, B_{\sigmaMeson  }  + \QQ{33} \, B_{G  }\, , \\
C_{\HField  } &= \QQ{21} \, C_{\sigmaMeson  } \, , \\
C_{\SField  } &= \QQ{22} \, C_{\sigmaMeson  } \, , \\
C_{\GField  } &= \QQ{23} \, C_{\sigmaMeson  }  \, .
\end{align}
\endgroup
With these amplitudes the decay widths are given by
\begin{align}\label{eq:physicalDecayWidths2}
\Gamma_{\HField\rightarrow \pi\pi} &= 3 \dfrac{k_f( M_\HField,m_\pi,m_\pi)}{4 \pi 
 M_{\HField}^2} \left| -i \mathcal{M}_{\HField  } \right|^2 \, , \\
\Gamma_{\SField\rightarrow \pi\pi} &= 3 \dfrac{k_f( M_\SField,m_\pi,m_\pi)}{4 \pi 
 M_{\SField}^2} \left| -i \mathcal{M}_{\SField  } \right|^2 \, , \\
\Gamma_{\GField\rightarrow \pi\pi} &= 3 \dfrac{k_f( M_\GField,m_\pi,m_\pi)}{4 \pi 
 M_{\GField}^2} \left| -i \mathcal{M}_{\GField  } \right|^2 \, , \label{eq:physicalDecayWidthsEnd}
\end{align}
where
\begin{align}
    -i \mathcal{M}_{\HField  } &=  i \left( A_{\HField  } - B_{\HField  } \dfrac{  M_{\HField}^2 - 2 m_{\pi}^2}{2}  
    - C_{\HField  } m_{\pi}^2 \right) \, , \\
    -i \mathcal{M}_{\SField  } &=  i \left( A_{\SField  } - B_{\SField  } \dfrac{  M_{\SField}^2 - 2 m_{\pi}^2}{2}  
    - C_{\SField  } m_{\pi}^2 \right) \, , \\
    -i \mathcal{M}_{\GField  } &=  i \left( A_{\GField  } - B_{\GField  } \dfrac{  M_{\GField}^2 - 2 m_{\pi}^2}{2}  
    - C_{\GField  } m_{\pi}^2 \right) \, .
\end{align}

\subsection{Pion-pion scattering} 

For the pion-pion scattering amplitude, we simply need to add  
the corresponding expression for the scalar glueball exchange to Eq.\ \eqref{eq:mHmS}. 
Then, the dimensionless scattering lengths are given as
\begin{widetext}

\begin{align}
m_\pi a_0^0 &= \dfrac{1}{32 \pi} \Bigg \lbrace 12 (g_1^2 - h_3) Z^4 w^2 m_{\pi}^2 - 10 \left( \lambda_1 
+ \dfrac{\lambda_2}{2} \right) Z^4 - 2 ( h + h_3) Z^4 w^2 m_{\pi}^2 \notag \\
&+ \dfrac{12}{\massH^2 - 4 m_\pi^2} \left[ (B_{\HField  } + C_{\HField  } ) m_{\pi}^2 - A_{\HField  } \right]^2 
+ \dfrac{12}{\massS^2 - 4 m_\pi^2} \left[ (B_{\SField  } + C_{\SField  } ) m_{\pi}^2 - A_{\SField  } \right]^2 \notag\\
&+  \dfrac{12}{\massG^2 - 4 m_\pi^2} \left[ (B_{\GField  } + C_{\GField  } ) m_{\pi}^2 - A_{\GField  } \right]^2 
+ \dfrac{8}{\massH^2} \left[ (B_{\HField  } - C_{\HField  } ) m_{\pi}^2 + A_{\HField  } \right]^2 \notag \\
&+  \dfrac{8}{\massS^2} \left[ (B_{\SField  } - C_{\SField  } ) m_{\pi}^2 + A_{\SField  } \right]^2 
+ \dfrac{8}{\massG^2} \left[ (B_{\GField  } - C_{\GField  } ) m_{\pi}^2 + A_{\GField  } \right]^2 
+ 16 g_1^2 Z^4 m_{\pi}^2 \dfrac{m_\rho^2}{m_{a_1}^4}\Bigg \rbrace \, ,
\end{align}
and
\begin{align}
m_\pi a_0^2 &= \dfrac{1}{16 \pi} \Bigg \lbrace -2 \left( \lambda_1 + \dfrac{\lambda_2}{2} \right) Z^4 
+ 2 ( h+ h_3) Z^4 w^2 m_{\pi}^2 - 4 g_1^2 Z^4 m_\pi^2 \dfrac{m_\rho^2}{m_{a_1}^4} \notag \\
&+ \dfrac{4}{\massH^2} \left[ (B_{\HField  } - C_{\HField  } ) m_{\pi}^2 + A_{\HField  } \right]^2 
+ \dfrac{4}{\massS^2} \left[ (B_{\SField  } - C_{\SField  } ) m_{\pi}^2 + A_{\SField  } \right]^2 \notag \\
& \dfrac{4}{\massG^2} \left[ (B_{\GField  } - C_{\GField  } ) m_{\pi}^2 + A_{\GField  } \right]^2 \Bigg \rbrace \, .
\end{align}

\end{widetext}

\pagebreak
\section{Pion-nucleon scattering parameters}
\label{app:piNScatFormulas}

\begin{figure}[ht]

\includegraphics{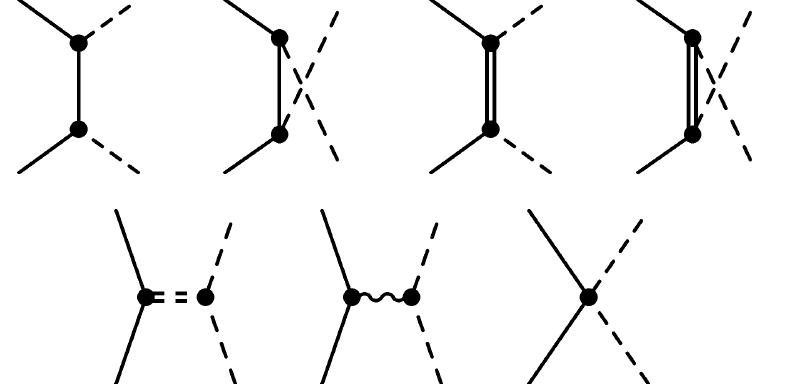}

\caption{Pion-nucleon scattering diagrams at tree level. The solid
line represents the nucleon, 
the double line its chiral partner, the dashed line the pion, the double-dashed the scalar-isoscalars 
$\HField$, $\SField$, and $\GField$, and the wavy line the $\rho$ meson.}
\label{fig:treeLevelDiagrams}
\end{figure}

The Feynman diagrams which contribute to the pion-nucleon scattering amplitude are shown in Fig.\ \ref{fig:treeLevelDiagrams}. The isospin-even and isospin-odd scattering lengths, scattering volumes, and effective range parameters can be calculated using \cite{Osypowski}:
\begin{align}
& a_{0}^{(\pm)} = \eta \left( A_{0}^{(\pm)} + m_{\pi} B_{0}^{(\pm)} \right) \, , \\
& a_{1 +}^{(\pm)} = \dfrac{2}{3} \eta C_{0}^{(\pm)}\, , \\
& a_{1 -}^{(\pm)} = \dfrac{2}{3} \eta C_{0}^{(\pm)} - \dfrac{\eta}{4 m_{N}^{2}} 
\left[ A_{0}^{(\pm)} - (2 m_{N} + m_{\pi}) B_{0}^{(\pm)} \right]\, , \\
& r_{0}^{(\pm)} = \eta \Biggl\lbrace - 2 C_{0}^{(\pm)} + \dfrac{(m_{N} 
+ m_{\pi})^{2}}{m_{N} m_{\pi}} D_{0}^{(\pm)} - \dfrac{1}{2 m_{N} m_{\pi}} \notag\\
& \, \, \, \, \, \, \, \times \left[ \left( 1 -\dfrac{m_{\pi}}{2 m_{N}} \right) A_{0}^{(\pm)} 
- \left(m_{N} + \dfrac{m_{\pi}^{2}}{2 m_{N}}\right) B_{0}^{(\pm)} \right]  \Biggr\rbrace\, ,
\end{align}
where 
\begin{align}
\eta & = \dfrac{1}{4 \pi \left( 1 + \dfrac{m_{\pi}}{m_{N}} \right)}\, , \\ 
C_{0}^{(\pm)} & = \dfrac{\partial}{\partial t} (A^{(\pm)} + m_{\pi} B^{\pm}) \big|_{t = 0} \, , \\ 
D_{0}^{(\pm)} & = \dfrac{\partial}{\partial s} (A^{(\pm)} + m_{\pi} B^{\pm}) \big|_{t = 0} \, .
\end{align}

The partial amplitudes $A^{(\pm)}$ and $B^{(\pm)}$ can be extracted by rewriting the scattering amplitude into the form
\begin{align}\label{eq:scatteringAmplitude}
T_{ab} &= \left[ A^{(+)} + \dfrac{(\slashed{q}_1+\slashed{q}_2)}{2}B^{(+)} \right] \delta_{ab} \notag \\
&+ \left[ A^{(-)} + \dfrac{ (\slashed{q}_1 + \slashed{q}_2)}{2}B^{(-)} \right] i \epsilon_{bac} \tau_c \, ,
\end{align}
were $a$ and $b$ are the isospin-indices of the inital and final pion states.

 The correct result for the partial amplitudes for the 
model of Ref.\ \cite{Gallas} reads:
\begin{widetext}
\begin{align}
A^{(+)} &=  4 \left(  g_{\pi N} \, g_{\partial \pi N} +  \,g_{\partial \pi N}^{2} \, m_{N}  
- \, g_{N^{*} \pi} g_{N^{*} \partial \pi} \right) - 2 g_{N^{*} \partial \pi}^{2} \, (m_{N^{*}} - m_{N}) + 2 g_{NN\pi\pi}  \notag\\
& + \left\lbrace - g_{N^{*} \pi}^{2} (m_{N} + m_{N^{*}}) + \, (m_{N}^{2} - m_{N^{*}}^{2}) \left[ 2 \, g_{N^{*} \pi} g_{N^{*} 
\partial \pi}  + g_{N^{*} \partial \pi}^{2} (m_{N^{*}} - m_{N}) \right] \right\rbrace \times  \left( \dfrac{1}{s - m_{N^{*}}^{2}} 
+ \dfrac{1}{u - m_{N^{*}}^{2}} \right) \notag\\
&-  \dfrac{2 \, g_{N \sigma}}{t - m_{\sigma}^{2}}\left[g_{\pi \sigma} + g_{\partial \pi \sigma} \left(m_{\pi}^{2} 
- \dfrac{t}{2}\right) + g_{\partial \sigma \pi} \dfrac{t}{2} \right] \, \\
A^{(-)} & =  \left\lbrace - g_{N^{*} \pi}^{2} (m_{N} + m_{N^{*}}) + (m_{N}^{2} - m_{N^{*}}^{2}) \left[ 2 \, g_{N^{*} \pi} 
g_{N^{*} \partial \pi}  + g_{N^{*} \partial \pi}^{2} (m_{N^{*}} - m_{N})\right] \right\rbrace \left( \dfrac{1}{s - m_{N^{*}}^{2}} 
- \dfrac{1}{u - m_{N^{*}}^{2}} \right) \, ,\notag\\
B^{(+)} & = -\left( g_{\pi N}  + 2 \, g_{\partial \pi N} \, m_{N} \right)^2 \left( \dfrac{1}{s - m_{N}^{2}}
 - \dfrac{1}{u - m_{N}^{2}} \right) - \left[ g_{N^{*} \pi} - g_{N^{*} \partial \pi} (m_{N^{*}} - m_{N})\right]^2 
 \left( \dfrac{1}{s - m_{N^{*}}^{2}} - \dfrac{1}{u - m_{N^{*}}^{2}} \right)\, ,\\
B^{(-)} & = - \left( g_{\pi N} +2  g_{\partial \pi N} \, m_{N} \right)^2 \left( \dfrac{1}{s - m_{N}^{2}} 
+ \dfrac{1}{u - m_{N}^{2}} \right) - \left[ g_{N^{*} \pi}  + g_{N^{*} \partial \pi} (m_{N^{*}} - m_{N}) \right]^2 
\left( \dfrac{1}{s - m_{N^{*}}^{2}} + \dfrac{1}{u - m_{N^{*}}^{2}} \right) \notag\\
&  - 2 \, g_{\partial \pi N}^{2} - 2 \, g_{N^{*} \partial \pi}^{2} + \dfrac{2 \, g_{N \rho}}{t - m_{\rho}^{2}} 
\left(g_{\pi \rho} + g_{\partial \pi \partial \rho} \dfrac{t}{2} \right) \, .
\end{align}
The couplings can be read off from the meson and the baryon Lagrangian:
\begingroup
\allowdisplaybreaks
\begin{align}
g_{\pi \rho} & = - g_{1} Z^{2} + g_{1}^{2} \varphi w Z^{2} - \varphi w Z^{2} \left[ m_{\rho}^{2} 
- m_{a_{1}}^{2} + (g_{1} \varphi)^{2} \right] \dfrac{1}{\varphi^{2}} = - g_{1} Z^{2} \dfrac{m_{\rho}^{2}}{m_{a_{1}}^{2}}  \, , \\
 g_{\partial \pi \sigma} & = g_{1} w Z^{2} (g_{1} \varphi w - 1 ) + \dfrac{ \varphi}{2} w^{2} Z^{2} (h_{1} + h_{2} - h_{3} ) \, , \\
 g_{\pi \sigma} & = - \varphi Z^{2}  \left(\lambda_{1} + \dfrac{\lambda_{2}}{2} \right) 
 = - \dfrac{Z}{2 f_{\pi}} \left(m_{\sigma}^{2} - \dfrac{m_{\pi}^{2}}{Z^{2}}\right) \, , \\
 g_{\partial \sigma \pi} & = g_{1} w Z^{2} \, , \\
 g_{\partial \pi \partial \rho} & = g_{2} Z^{2} w^{2} \, , \\
 g_{\chiMeson \pi} & = g_{\chiMeson} Z^2 \, , \\
 g_{\chiMeson \partial \pi} & = \gAV Z^2 w^2 \, ,\\
 g_{\pi N} &= - Z \dfrac{e^{\delta} \hat{g}_{1} + e^{-\delta} \hat{g}_{2} }{4 \, \textrm{cosh} \, \delta}  \, ,\\
 g_{\pi N N^{*}} &= - g_{\pi N^{*} N} = Z \dfrac{- \hat{g}_{1} + \hat{g}_{2}}{4 \, \textrm{cosh} \, \delta} \, ,\\
 g_{\partial \pi N} &=  Z w \dfrac{e^{\delta} c_{1} - e^{- \delta} c_{2}}{4 \,  \textrm{cosh} \, \delta} \, , \\
 g_{\partial \pi N N^{*}} &= g_{\partial \pi N^{*} N} = - Z w \dfrac{c_{1} + c_{2}}{4 \, \textrm{cosh} \, \delta}  \, ,
\end{align}
From this we obtain the isospin-even and isospin-odd scattering parameters:
\begin{align}
a_{0}^{(+)} & = \, \dfrac{1}{4 \pi \left( 1 + \dfrac{m_{\pi}}{m_{N}} \right)} \left( \dfrac{Z}{2 \, \textrm{cosh} \, \delta} \right)^{2}  
 \Biggl( - \dfrac{1}{2} ( \hat{g}_{1} - \hat{g}_{2})^2 \left[ 1 - \frac{Zf_\pi}{2} w \, (c_{1} + c_{2})\right]^{2}  
 \dfrac{ (m_{N^{*}}+m_N) \, (m_{N}^{2} + m_{\pi}^{2} - m_{N^{*}}^{2})}{(m_{N}^{2} + m_{\pi}^{2} - m_{N^{*}}^{2})^{2} 
 - 4 m_{N}^{2} m_{\pi}^{2}}  \notag\\
&  - w \, (c_{1} + c_{2}) ( \hat{g}_{1} - \hat{g}_{2}) \left[1 - \dfrac{Z f_\pi}{4}w (c_{1} + c_{2})\right] - w \, (c_{1} e^{\delta} 
- c_{2} e^{- \delta})\left[ \hat{g}_{1} e^{\delta} + \hat{g}_{2} e^{- \delta}- w \, m_{N} \, (c_{1} e^{\delta} - c_{2} e^{- \delta})
\right]   \notag\\
&  + (\hat{g}_{1} e^{\delta} - \hat{g}_{2} e^{- \delta}) \dfrac{\textrm{cosh} \, \delta}{Z f_{\pi}} \left\lbrace 1 
+ \dfrac{m_{\pi}^{2}}{m_{\sigma}^{2} Z^{4}} \left[ Z^{2} - 2 - 2(Z^{2} - 1)
\left( 1 - \dfrac{Z^{2} m_{1}^{2}}{m_{a_{1}}^{2}} \right) \right] \right\rbrace \notag\\
& + m_{\pi} \bigg\lbrace (\hat{g}_{1} - \hat{g}_{2})^2 \left[ 1 - \frac{Zf_\pi}{2} w \, (c_{1} + c_{2}) \right]^{2} 
\dfrac{m_{N}m_{\pi} }{(m_{N}^{2} + m_{\pi}^{2} - m_{N^{*}}^{2})^{2} - 4 m_{N}^{2} m_{\pi}^{2}} \notag\\
& \hspace*{0.7cm} + \left[\hat{g}_{1} e^{\delta} + \hat{g}_{2} e^{-\delta} 
- 2 \, w \, m_{N} \, (c_{1} e^{\delta} - c_{2} e^{-\delta})  \right]^{2} \dfrac{m_{N}}{ m_{\pi}} 
\dfrac{1}{m_{\pi}^{2} - 4 m_{N}^{2} } \bigg\rbrace
\Biggr) \, .
\end{align}
\endgroup
Notice the change of sign in front of the third term in brackets in the third line as compared to Eq.\ (18) of
Ref.\ \cite{Gallas}. The correct result for the isospin-odd scattering length reads
\begin{align}
a_{0}^{(-)} & = \, \dfrac{1}{4 \pi \left( 1 + \dfrac{m_{\pi}}{m_{N}} \right)} \left( \dfrac{Z}{2 \, \textrm{cosh} \, \delta} \right)^{2}  
\Biggl( (\hat{g}_{1} - \hat{g}_{2})^2 \left[1- \frac{Zf_\pi}{2} w \, (c_{1} + c_{2})\right]^{2}  \dfrac{ (m_{N} + m_{N^{*}})  
m_{N}m_{\pi}}{(m_{N}^{2} + m_{\pi}^{2} - m_{N^{*}}^{2})^{2} - 4 m_{N}^{2} m_{\pi}^{2}} \notag\\
& + \dfrac{ m_{\pi}}{2} \Biggl\lbrace - (\hat{g}_{1} - \hat{g}_{2} )^2 \left[1- \frac{Zf_\pi}{2} w \,(c_{1} + c_{2}) \right]^{2} 
\dfrac{m_{N}^{2} + m_{\pi}^{2} - m_{N^{*}}^{2}}{(m_{N}^{2} + m_{\pi}^{2} - m_{N^{*}}^{2})^{2} 
- 4 m_{N}^{2} m_{\pi}^{2}} \notag\\
&\hspace*{1.2cm} -  \bigl[ \hat{g}_{1} e^{\delta} + \hat{g}_{2} e^{-\delta} - 2 w \, m_{N} (c_{1} e^{\delta} - c_{2} e^{-\delta})  
\bigl]^{2} \dfrac{1}{m_{\pi}^{2} - 4 m_{N}^{2}} - w^{2} \left[ (c_{1} + c_{2})^{2} +  (c_{1} e^{\delta} 
- c_{2} e^{-\delta})^{2}\right] \notag\\
& \hspace*{1.2cm}
+ 4 \, \textrm{cosh} \, \delta \, \dfrac{ g_{1}}{m_{a_{1}}^{2}}\, (c_{1} e^{\delta} + c_{2} e^{-\delta})  \Biggr\rbrace  \Biggr)\;.
\end{align}
\end{widetext}
Errors in Eq.\ (19) of Ref.\ \cite{Gallas} were: (i) the sign of the first term in braces, (ii) the sign in front of the
second term in the second set of brackets in the third line, (iii) the coefficient of the last term ($1/m_{a_1}^2$ 
instead of $1/Z^2 m_\rho^2$), and (iv) the sign in front of the last term in parentheses in the fourth line.

In order to obtain the expressions for the scattering parameters for our model with scalar four-quark state
and dynamical scalar glueball we need to modify the pion-nucleon scattering amplitudes by 
replacing the last term of $A^{(+)}$ with the expression
\begin{align}
& - \dfrac{2 \, g_{N \HField}}{t - \massH^{2}}\left[g_{\pi \HField} + g_{\partial \pi \HField} \left(m_{\pi}^{2} - \dfrac{t}{2}\right) 
+ g_{\partial \HField \pi} \dfrac{t}{2} \right] \notag \\
&- \dfrac{2 \, g_{N \SField}}{t - \massS^{2}}\left[g_{\pi \SField} + g_{\partial \pi \SField} \left(m_{\pi}^{2} - \dfrac{t}{2}\right) 
+ g_{\partial \SField \pi} \dfrac{t}{2} \right] \, \notag\\
&- \dfrac{2 \, g_{N \GField}}{t - \massG^{2}}\left[g_{\pi \GField} + g_{\partial \pi \GField} \left(m_{\pi}^{2} - \dfrac{t}{2}\right) 
+ g_{\partial \GField \pi} \dfrac{t}{2} \right] \, ,
\end{align}
where
\begingroup
\allowdisplaybreaks
\begin{align}
g_{N \HField} &= \QQ{11} \, g_{N \chiMeson} + \QQ{21} \, g_{N \sigmaMeson} + \QQ{31} \, g_{N G} \, , \\
g_{N \SField} &= \QQ{12} \, g_{N \chiMeson} + \QQ{22} \, g_{N \sigmaMeson} + \QQ{32} \, g_{N G} \, , \\
g_{N \GField} &= \QQ{13} \, g_{N \chiMeson} + \QQ{23} \, g_{N \sigmaMeson} + \QQ{33} \, g_{N G} \, , \\
g_{\pi \HField} &= \QQ{11} \, g_{\pi \chiMeson} + \QQ{21} \,  g_{\pi \sigmaMeson} + \QQ{31} \, g_{\pi G}\,  , \\
g_{\pi \SField} &= \QQ{12} \, g_{\pi \chiMeson} + \QQ{22} \,  g_{\pi \sigmaMeson} + \QQ{32} \, g_{\pi G}\,  , \\
g_{\pi \GField} &= \QQ{13} \, g_{\pi \chiMeson} + \QQ{23} \,  g_{\pi \sigmaMeson} + \QQ{33} \, g_{\pi G}\,  , \\
g_{\partial \pi \HField} &= \QQ{11} \, g_{\partial \pi \chiMeson} + \QQ{21} \,g_{\partial \pi \sigmaMeson} 
+ \QQ{31} \,g_{\partial \pi G}  \, , \\
g_{\partial \pi \SField} &= \QQ{12} \, g_{\partial \pi \chiMeson} + \QQ{22} \,g_{\partial \pi \sigmaMeson} 
+ \QQ{32} \,g_{\partial \pi G}  \, , \\
g_{\partial \pi \GField} &= \QQ{13} \, g_{\partial \pi \chiMeson} + \QQ{23} \,g_{\partial \pi \sigmaMeson} 
+ \QQ{33} \,g_{\partial \pi G}  \, , \\
g_{\partial \HField \pi} &= \QQ{11} \, g_{\partial \sigmaMeson \pi} \, , \\
g_{\partial \SField \pi} &= \QQ{12} \, g_{\partial \sigmaMeson \pi} \, , \\
g_{\partial \GField \pi} &= \QQ{13} \, g_{\partial \sigmaMeson \pi} \, .
\end{align}
\endgroup
and
\begin{align}
    g_{N G} &= - \dfrac{b}{\cosh \, \delta} \, , \\
    g_{\pi G} &= - Z^2 \dfrac{\mu^2}{G_0} + Z^2 \dfrac{g_\chiMeson}{G_0} \chiMeson_0 \, ,\\
    g_{\partial \pi G} &= Z^2 w^2 \left( \dfrac{m_1^2}{G_0} + \dfrac{\gAV \chiMeson_0}{G_0} \right) \, .
\end{align}

\bibliography{paperRefs}{}
\end{document}